%% file: CTEQ66AS.tex
\documentclass[12pt,english,prd,nofootinbib,preprint,floatfix,tightenlines]{revtex4}

\usepackage{amsmath,amssymb,graphicx}
\usepackage{color}


\input CTEQ66AS-figures.tex     % figure and table definitions
\input CTEQ66AS-tables.tex     % figure and table definitions

\begin{document}

\newcommand{\DATE}  {\today}

\preprint{MSUHEP-100421, SMU-HEP-10-07, arXiv:1004.4624 [hep-ph]}
\title{Uncertainty induced by QCD coupling \\ in 
the CTEQ global analysis of parton distributions}

\author{Hung-Liang Lai,$^{1,2}$ Joey Huston,$^{2}$
Zhao Li,$^{2}$ Pavel Nadolsky,$^{3}$ \\
Jon Pumplin,$^{2}$, Daniel Stump,$^{2}$ and C.-P.\ Yuan$^{2}$}

\affiliation
{ $^{1}$Taipei Municipal University of Education, Taipei, Taiwan \\
$^{2}$Department of Physics and Astronomy, Michigan State University,
\\
East Lansing, MI 48824-1116, U.S.A.\\
$^{3}$ Department of Physics, Southern Methodist University, \\
Dallas, TX 75275-0175, U.S.A.}

\date{\today}

\begin{abstract}
We examine the dependence of parton distribution functions
(PDFs) on the value of the QCD coupling strength $\alpha_{s}(M_{Z})$.
We explain a simple method that is rigorously valid 
in the quadratic approximation normally applied in PDF 
fitting, and fully reproduces the correlated dependence of 
theoretical cross sections on $\alpha_s$
and PDF parameters.
This method is based on a
statistical relation that allows one to add the uncertainty produced by $\alpha_s$,
computed with some special PDF sets, in quadrature with the PDF
uncertainty obtained for the fixed $\alpha_s$ value (such as the
CTEQ6.6 PDF set). A series of four CTEQ6.6AS PDFs realizing this
approach, for $\alpha_s$ values in the interval 
$0.116 \leq \alpha_{s}(M_{Z}) \leq 0.120$, is presented.
Using these PDFs, the combined $\alpha_{s}$ and PDF uncertainty
is assessed for theoretical predictions at the Fermilab Tevatron 
and Large Hadron Collider. 
\end{abstract}



\maketitle
\tableofcontents{}

\newpage

\input CTEQ66AS-sec1.tex  % Section 1

\input CTEQ66AS-sec2.tex  % Section 2

\input CTEQ66AS-sec3.tex  % Section 3

\input CTEQ66AS-sec4.tex  % Section 4

\input CTEQ66AS-sec5.tex  % Section 5

\input CTEQ66AS-app1.tex  % Appendix with the proof

\input CTEQ66AS-cit.tex  % Bibliography

\end{document}

%% file: CTEQ66AS-figures.tex
\newcommand{\figGUSlowQ}
{
\begin{figure}[p]
\begin{center}$
\begin{array}{c}
\includegraphics[width=5.0in]{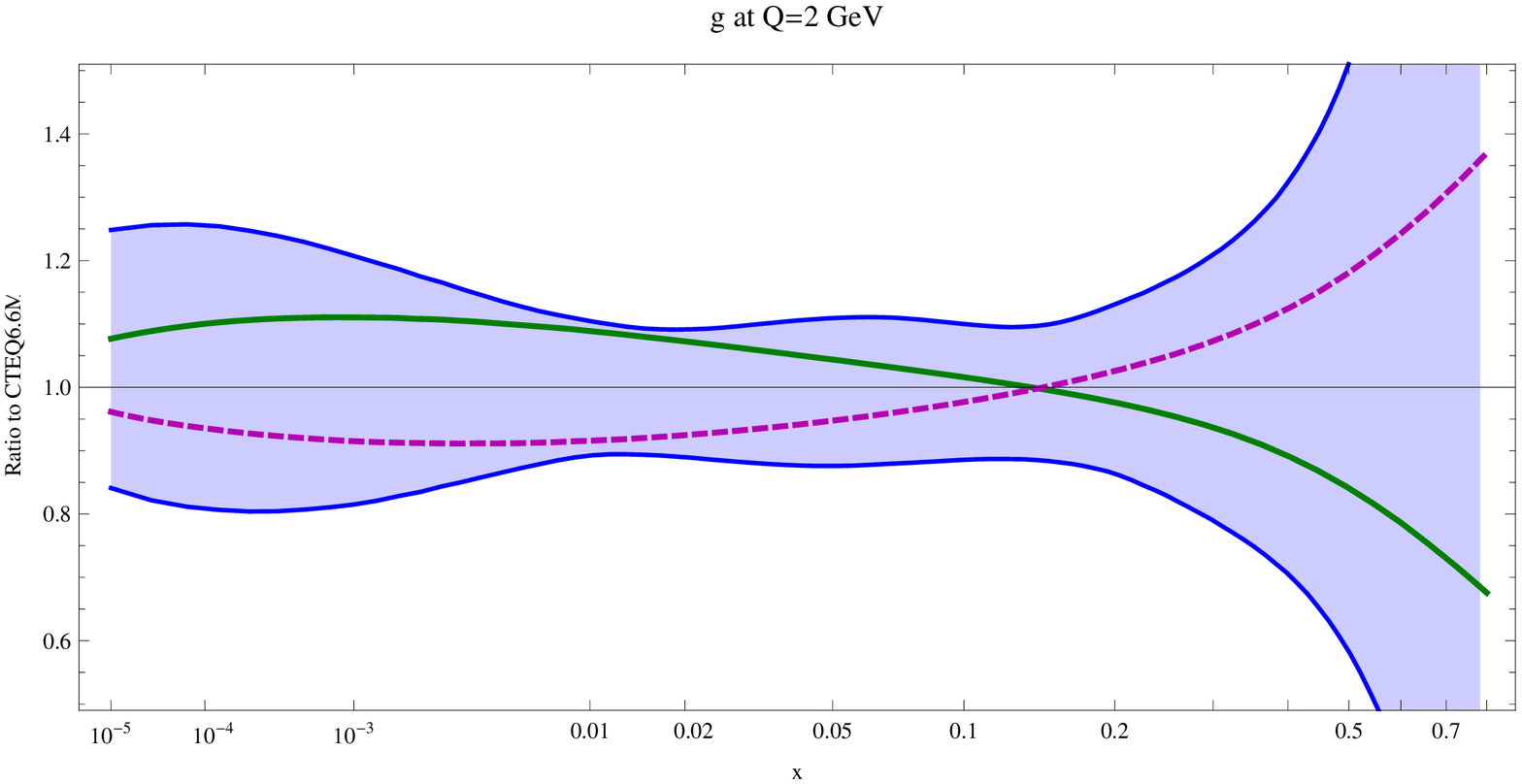} 
\\
\includegraphics[width=5.0in]{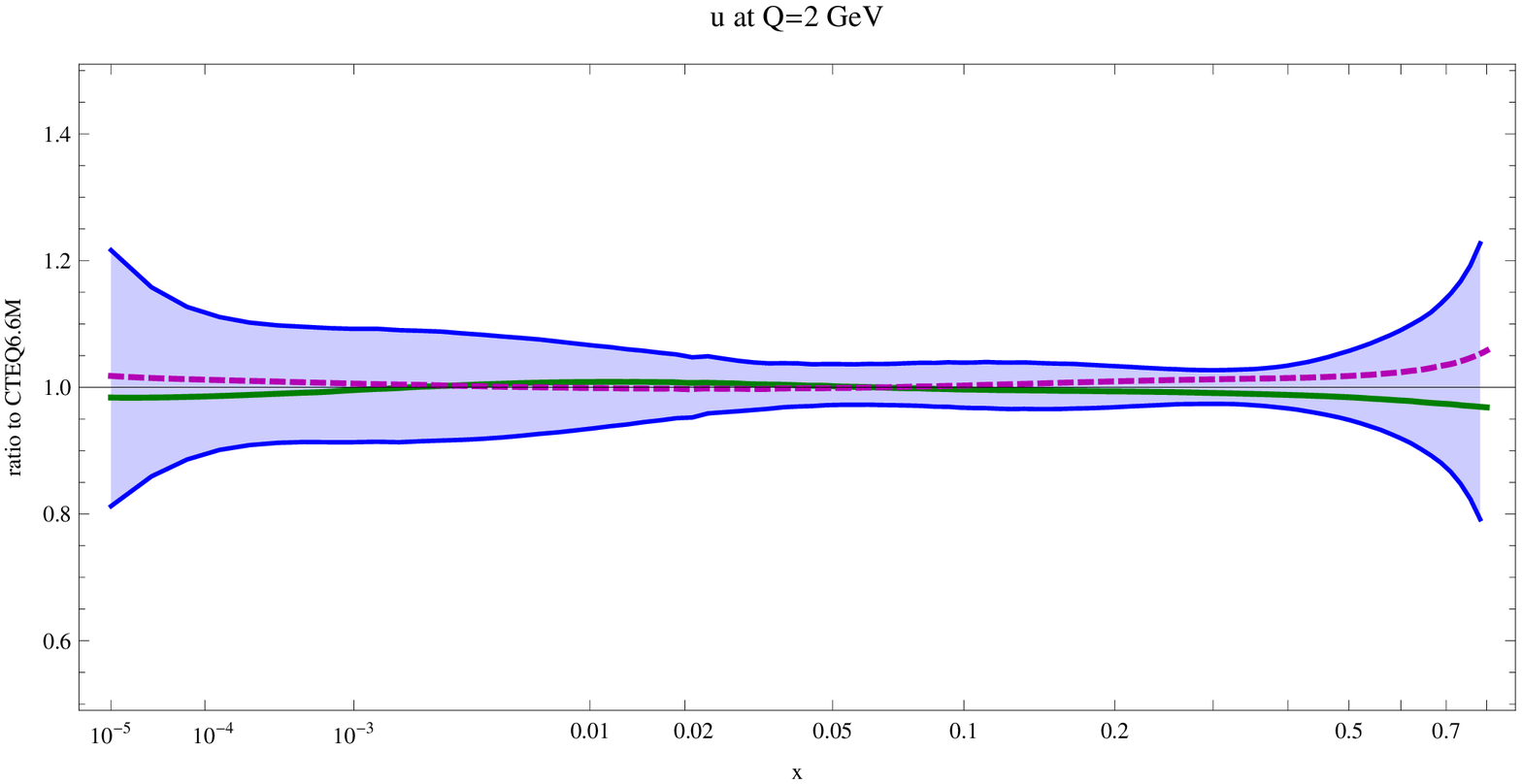}
\\
\includegraphics[width=5.0in]{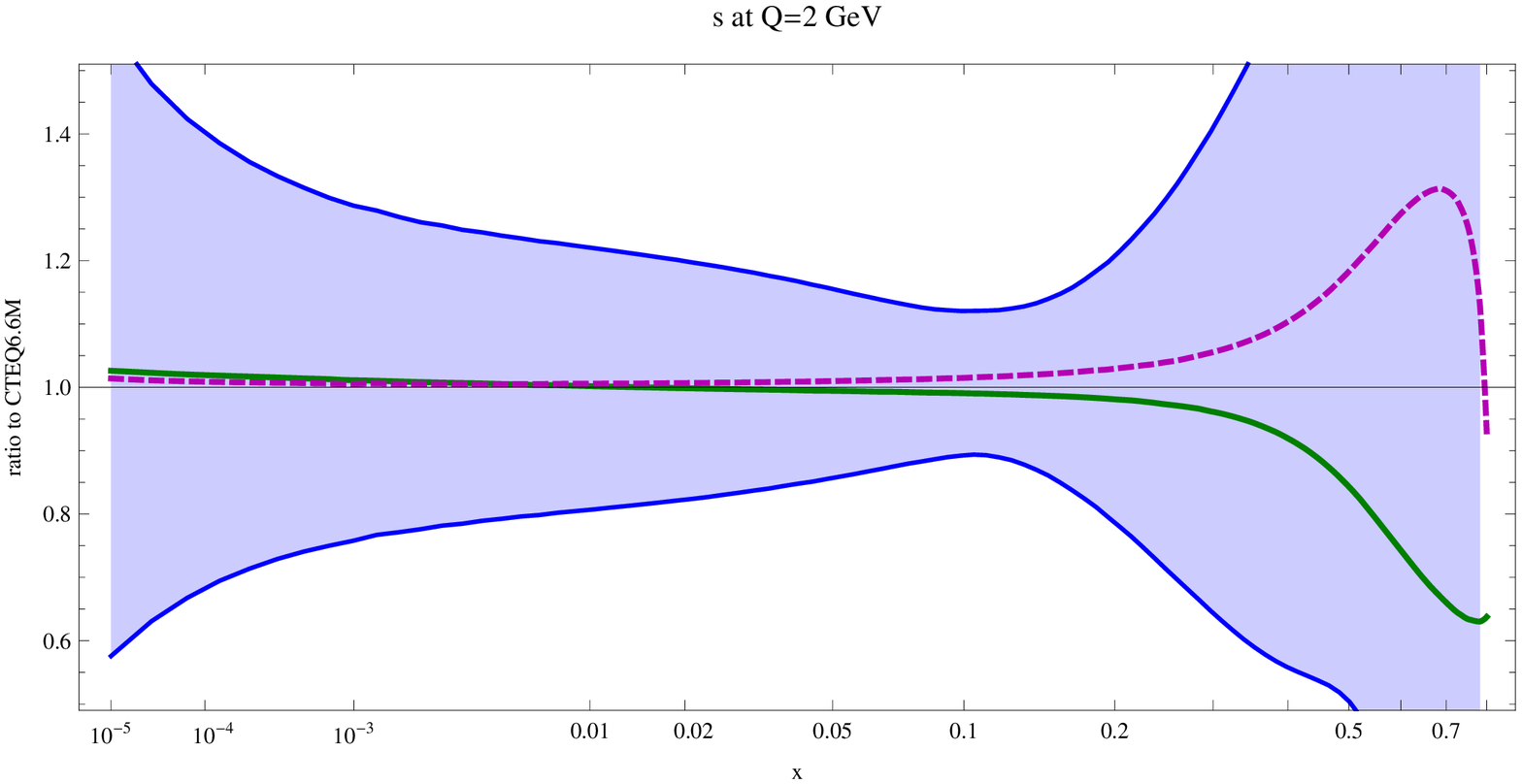}
\end{array}$
\end{center}
\caption{Comparison of the PDF and $\alpha_s$ uncertainties 
for the gluon, $u$ quark, and $s$ quark, at $Q = 2\,{\rm GeV}$.
The vertical axis is the {\em ratio} of $f(x,Q)$
to the CTEQ6.6M best fit.
The CTEQ6.6 PDF uncertainty range for the central $\alpha_{s}$
value, 0.118, is shown as the shaded region.
The ratio of $AS_{+2}$ and $AS_{-2}$ PDFs to the 
corresponding CTEQ6.6M PDFs are shown as the solid
and dashed curves, respectively.
}
\label{fig:GUSlowQ}
\end{figure}
}

\newcommand{\figGUShighQ}
{
\begin{figure}[p]
\begin{center}$
\begin{array}{c}
\includegraphics[width=5.0in]{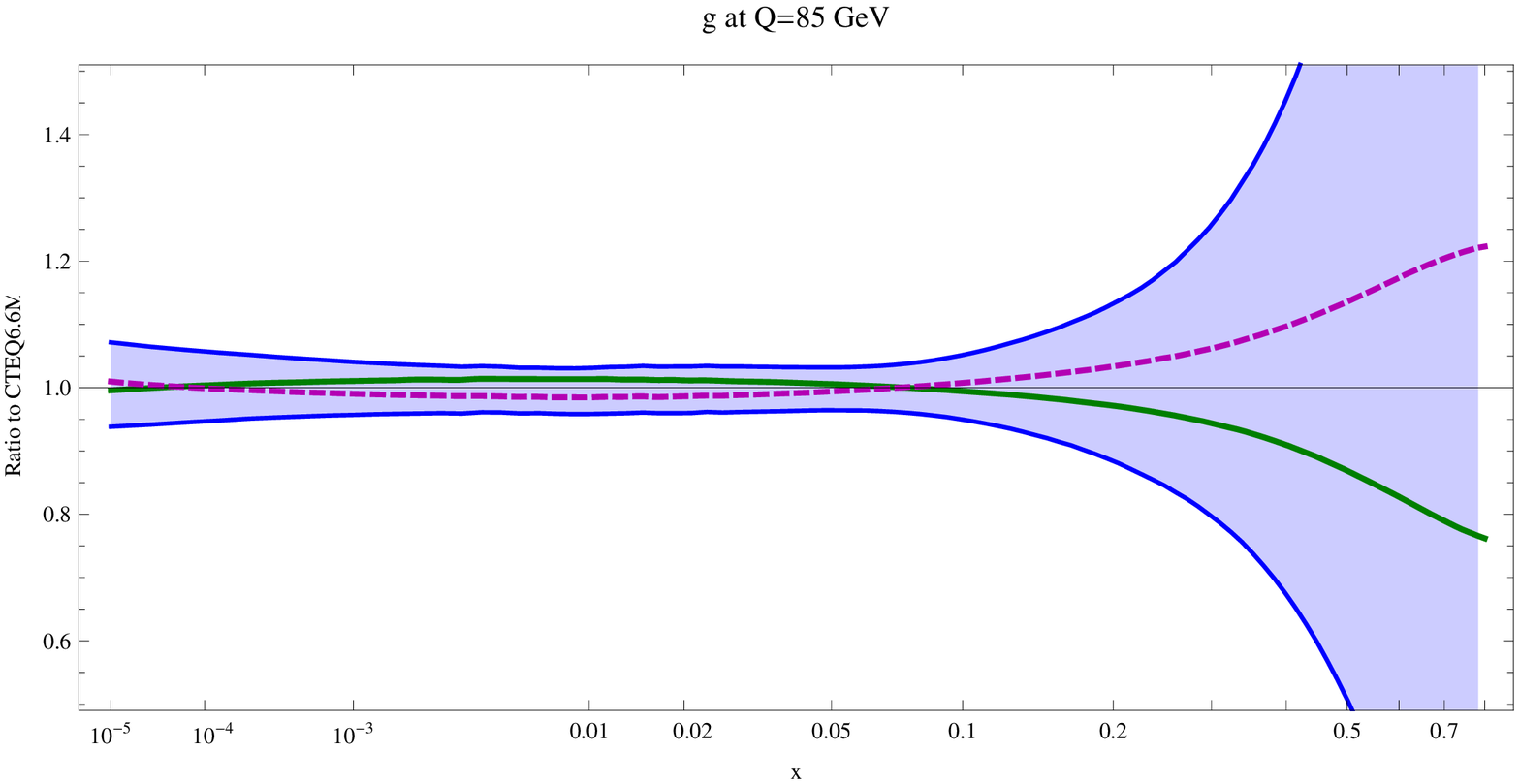} 
\\
\includegraphics[width=5.0in]{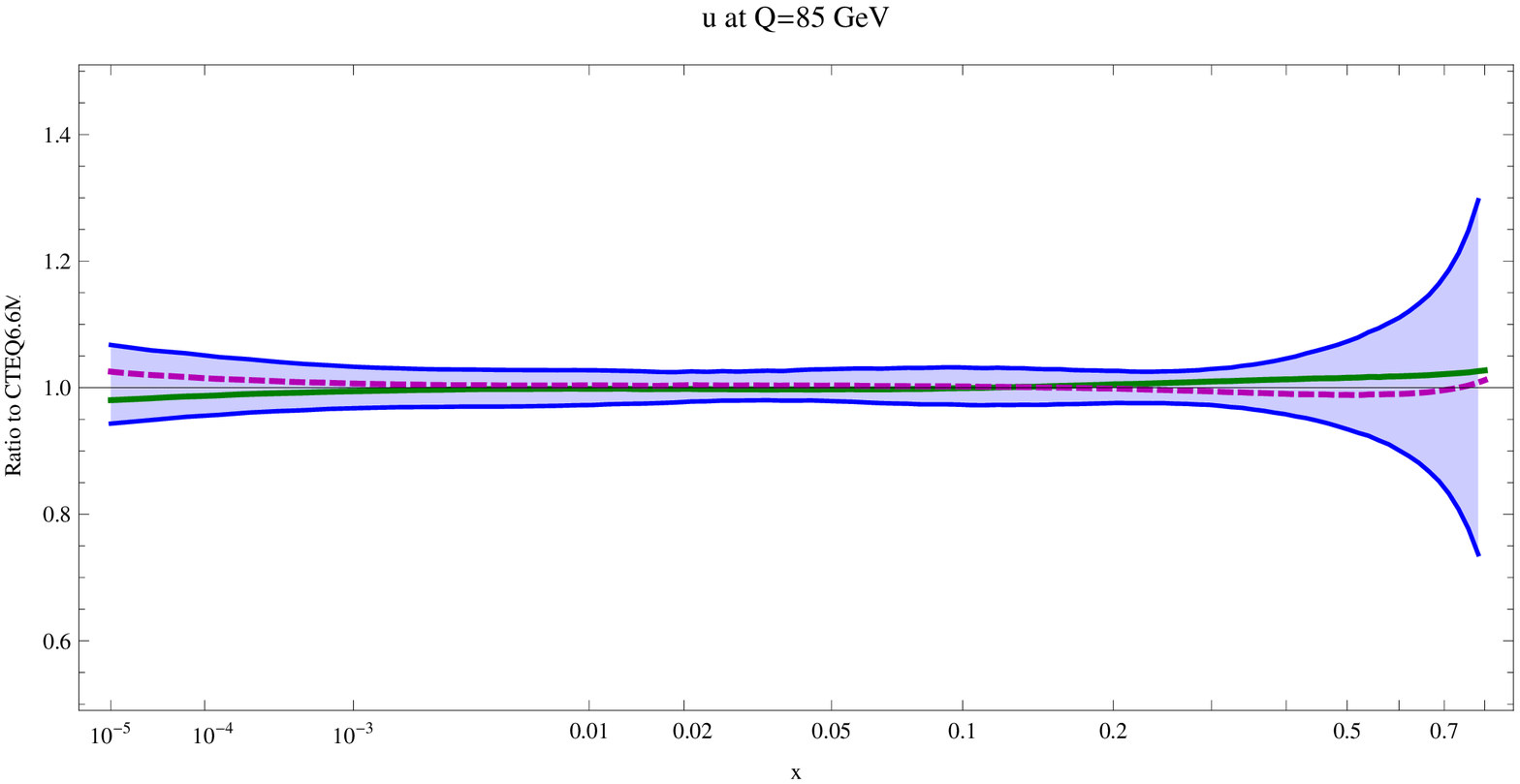}
\\
\includegraphics[width=5.0in]{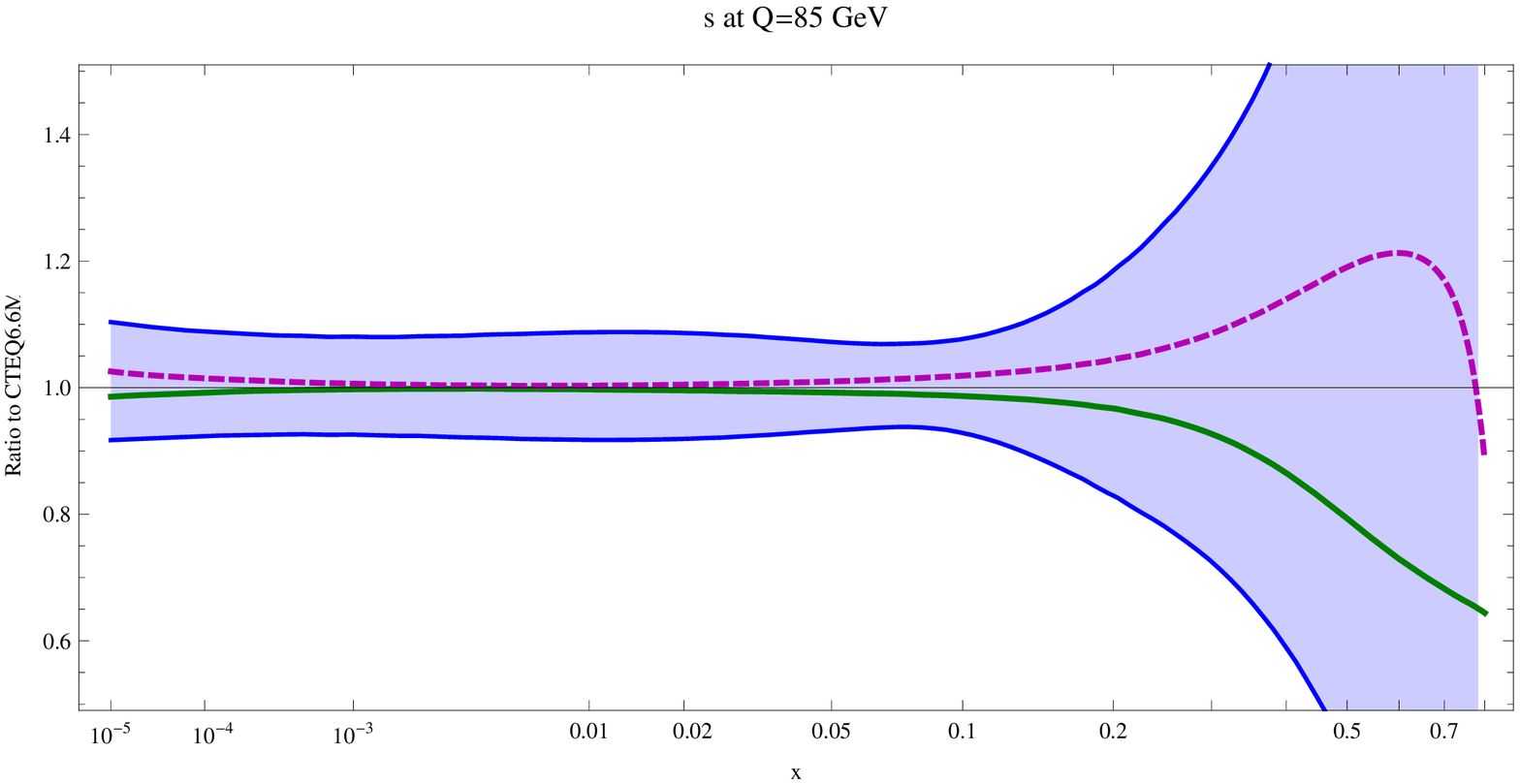}
\end{array}$
\end{center}
\caption{
The same as Fig.\,\ref{fig:GUSlowQ}, for $Q = 85$\,GeV.
}
\label{fig:GUShighQ}
\end{figure}
}

\newcommand{\figDRSttblhc}
{
\begin{figure}[tbh]
\begin{center}$
\begin{array}{c}
\includegraphics[width=5.0in]{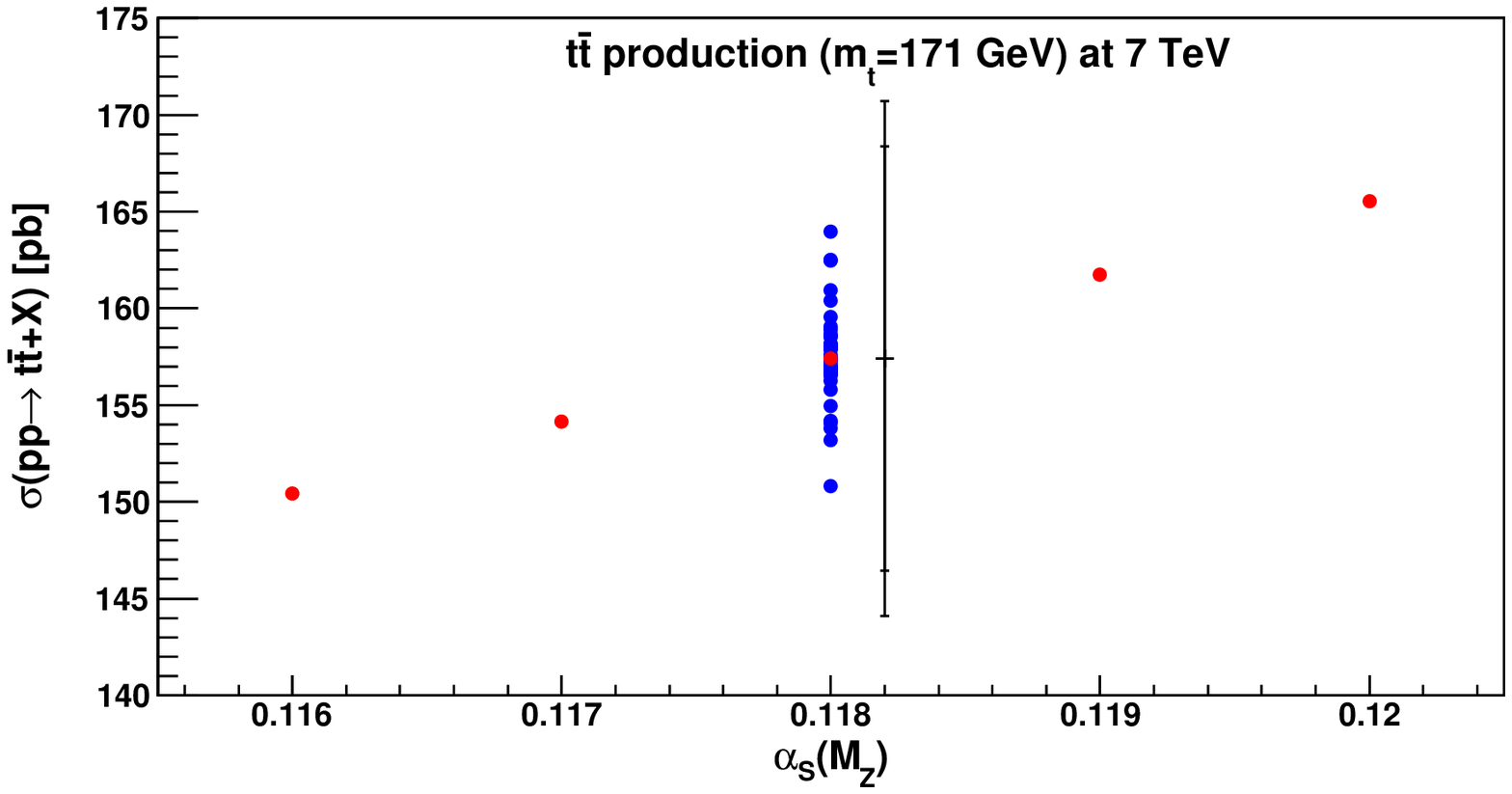} 
\\
\includegraphics[width=5.0in]{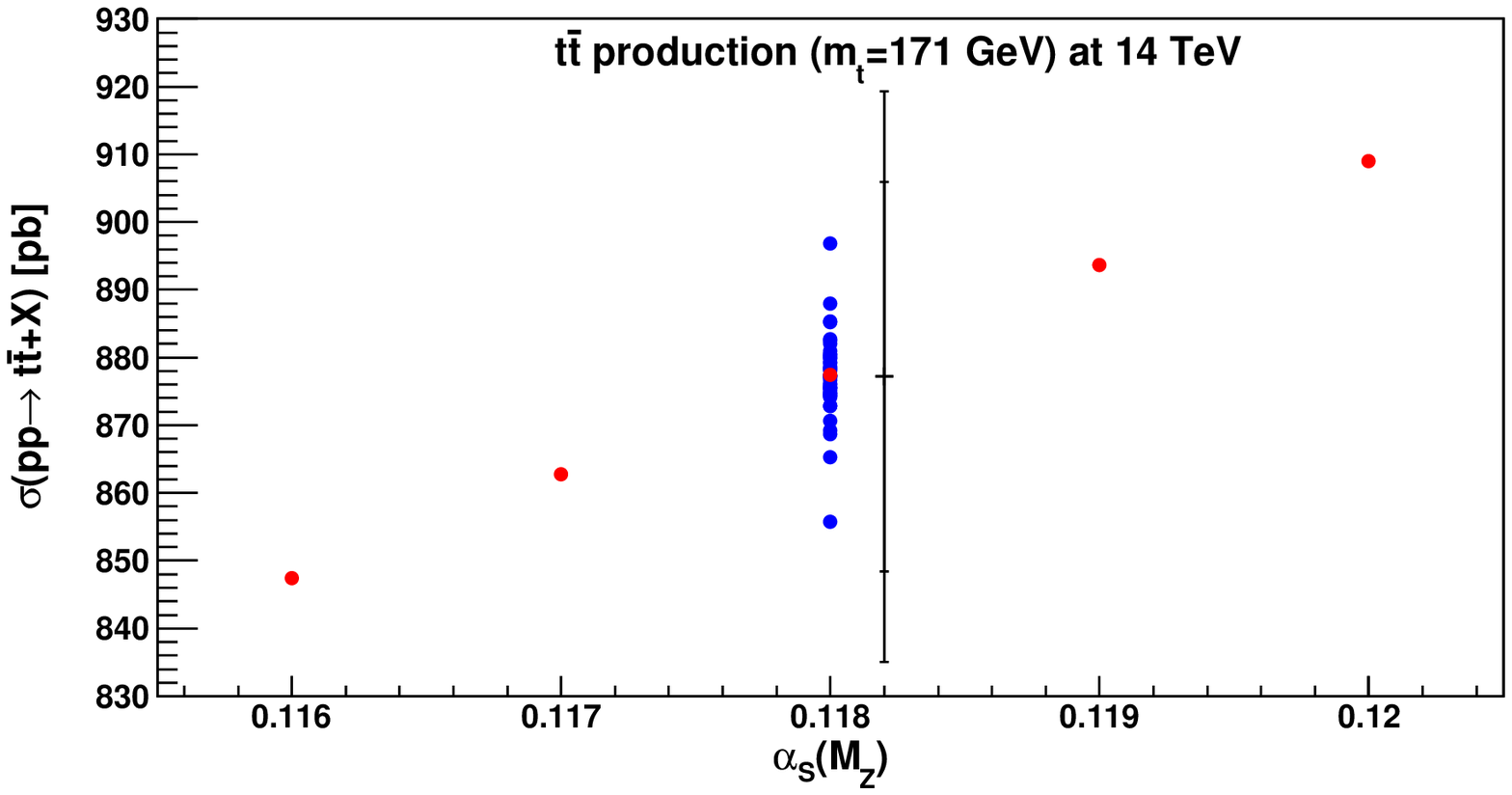}
\end{array}$
\end{center}
\caption{Cross section for $t\overline{t}$ production at the LHC,
at center-of-mass energies of 7 TeV and 14 TeV,
as a function of $\alpha_{s}(M_{Z})$.
The predictions for the the central set and the 44 eigenvector PDF  sets are shown
for $\alpha_{s}(M_{Z})=0.118$.
For the other values of $\alpha_{s}(M_{Z})$ (0.116, 0.117, 0.119, 0.120),
only the central prediction is shown.
The combined uncertainty range (CTEQ6.6+CTEQ6.6AS)
is shown as the error bar;
cf.\ Table \ref{tbl:zhaoli}.
The inner error bar is the PDF error alone,
and the outer error bar is the combined $\mathrm{PDF} \, + \, \alpha_s$ uncertainty.
}
\label{fig:DRSttblhc}
\end{figure}
}

\newcommand{\figDRSgghlhc}
{
\begin{figure}[tbh]
\begin{center}$
\begin{array}{c}
\includegraphics[width=5.0in]{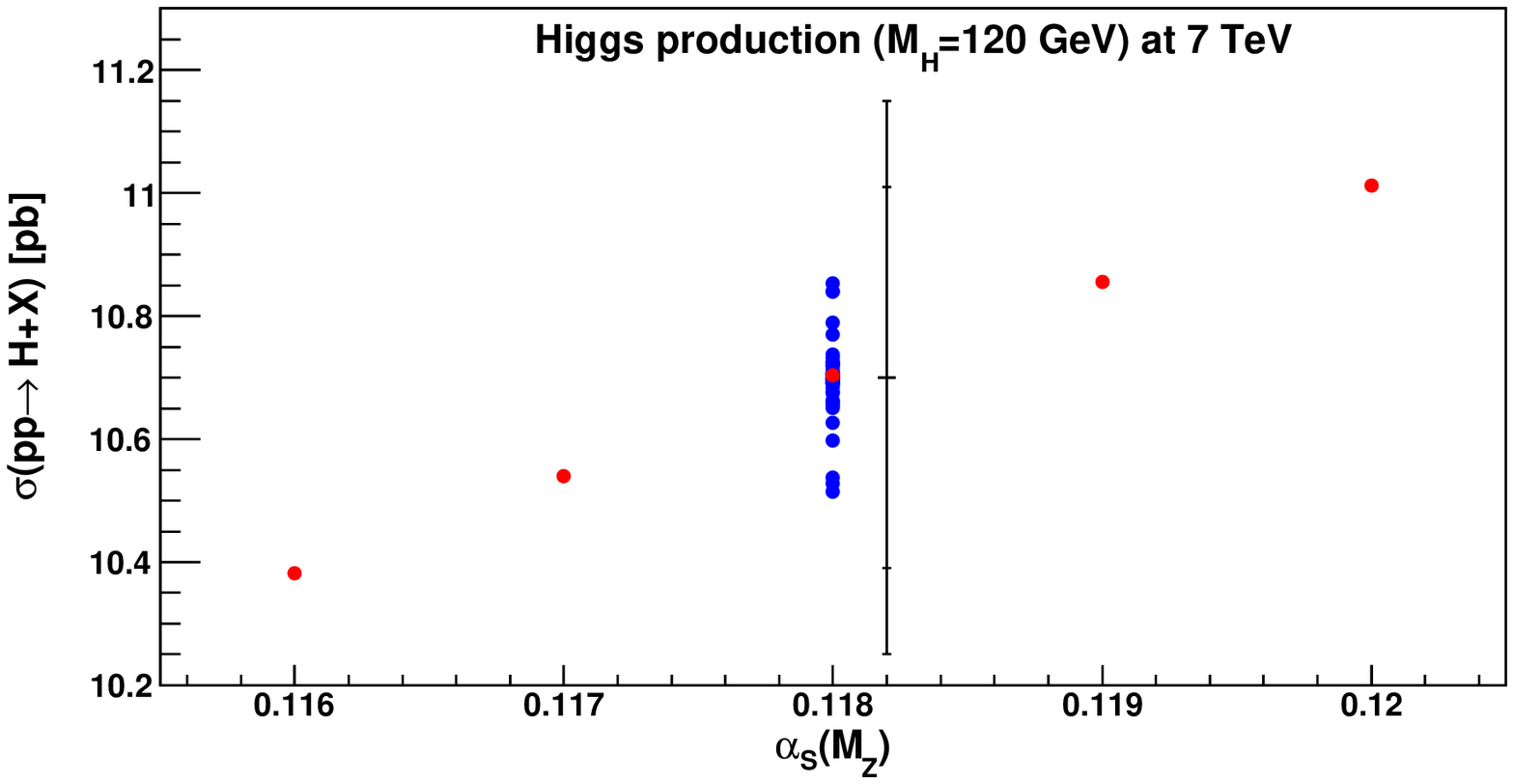} 
\\
\includegraphics[width=5.0in]{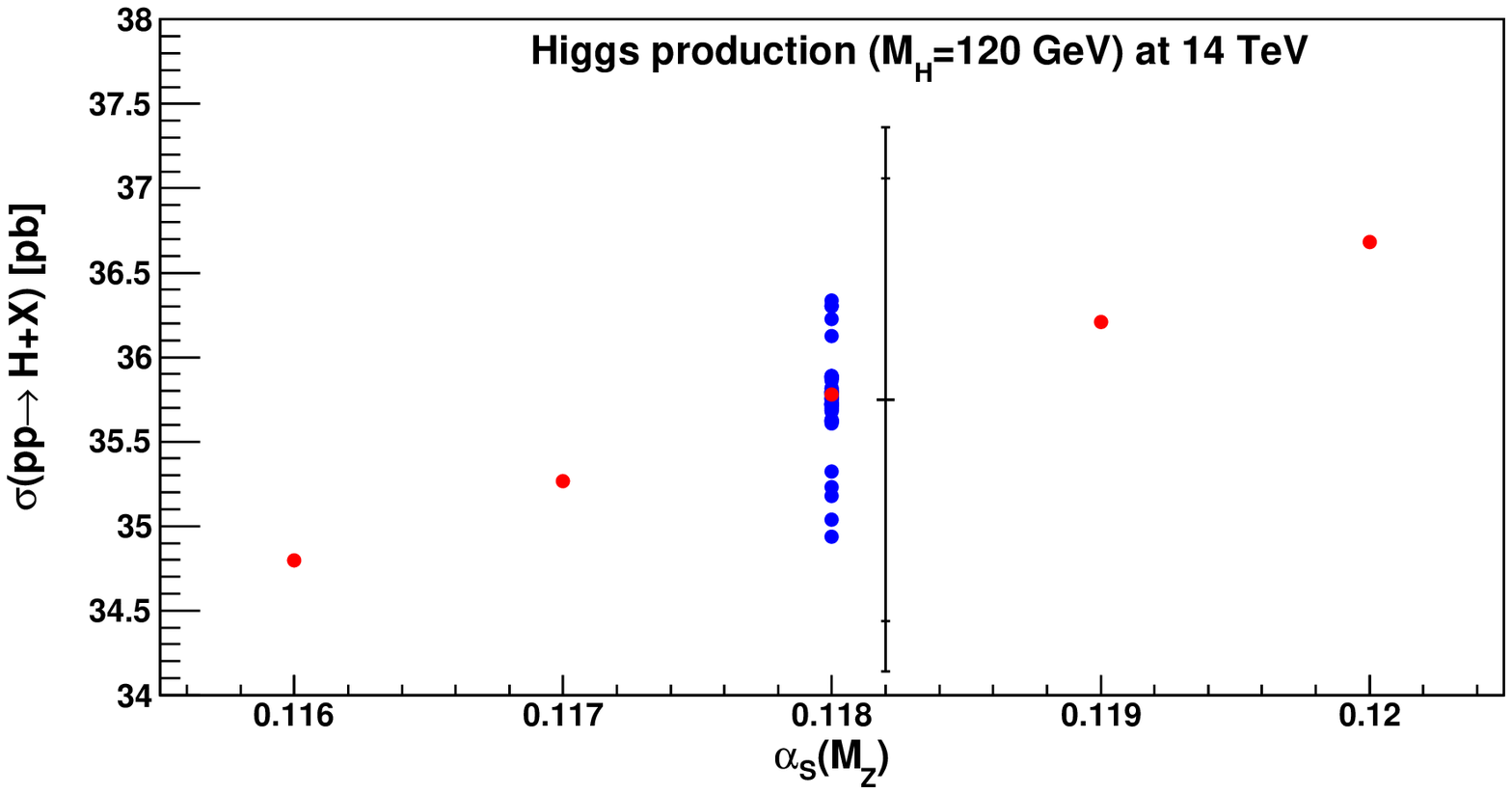}
\end{array}$
\end{center}
\caption{Same as Fig.\,\ref{fig:DRSttblhc}, for production
of Standard Model Higgs boson with mass 120 GeV at the LHC.
}
\label{fig:DRSgghlhc}
\end{figure}
}

\newcommand{\figPNbands}
{
\begin{figure}
\includegraphics[width=3.7in]{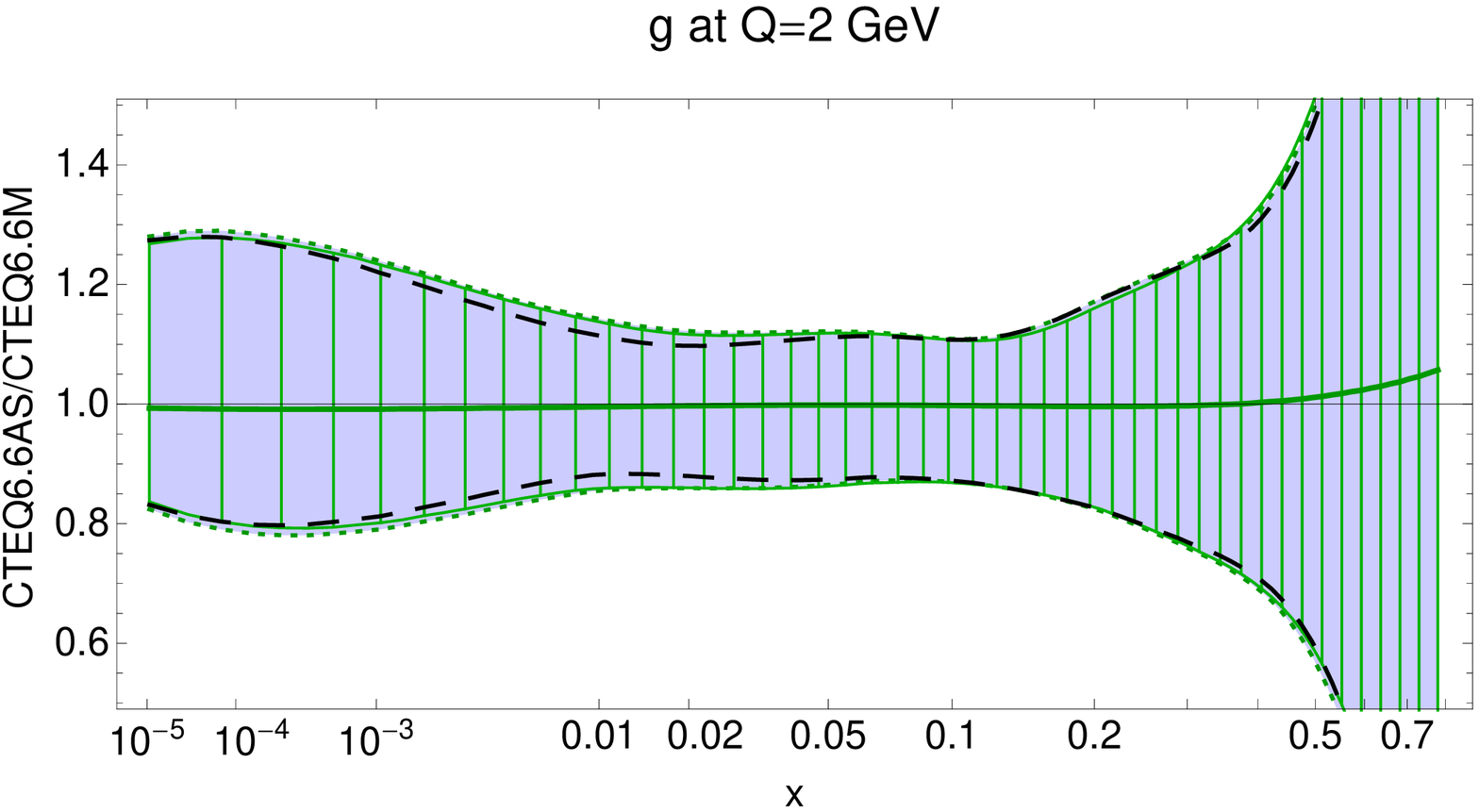}
~~\includegraphics[width=3.7in]{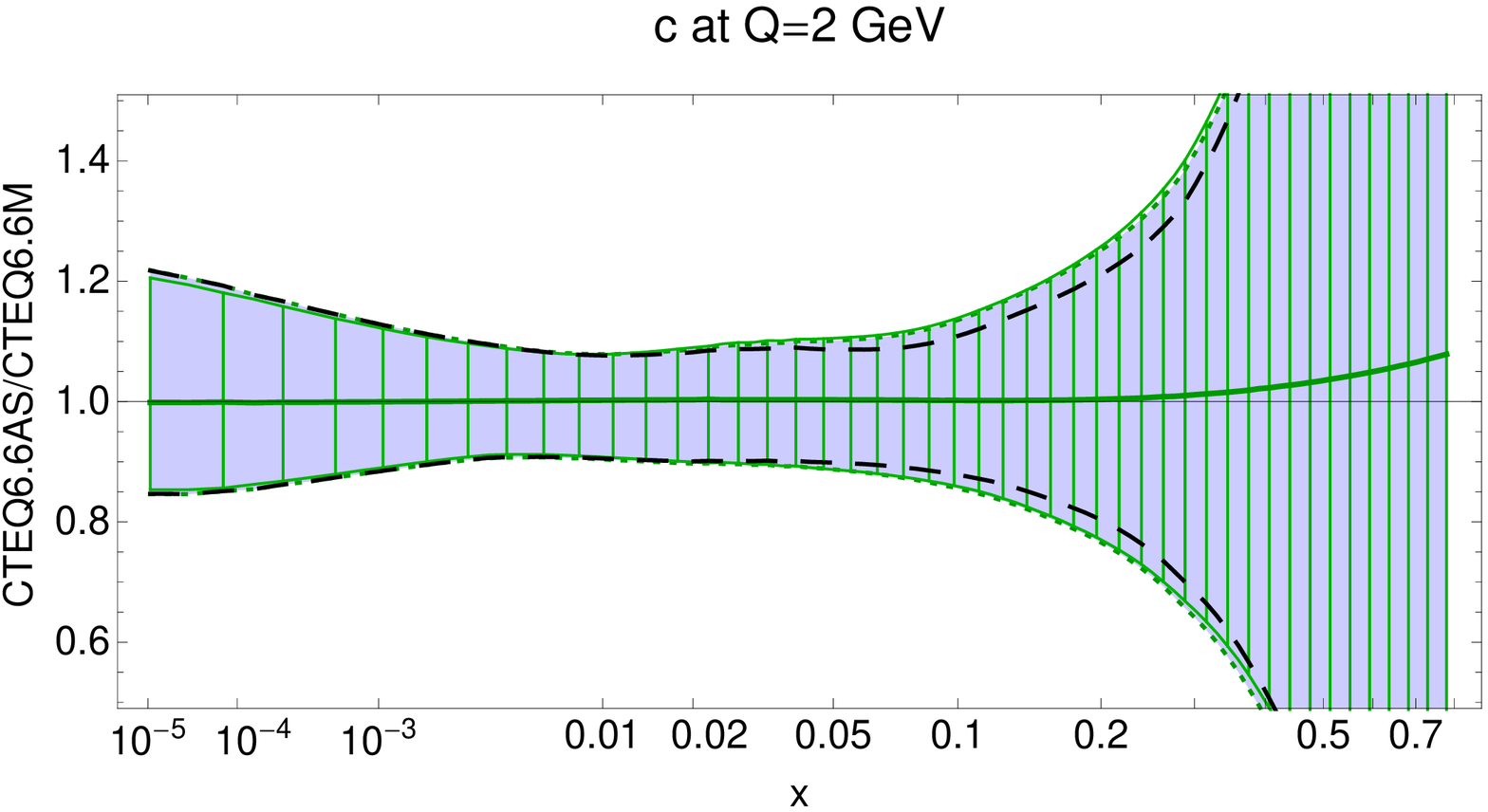}

\includegraphics[width=3.7in]{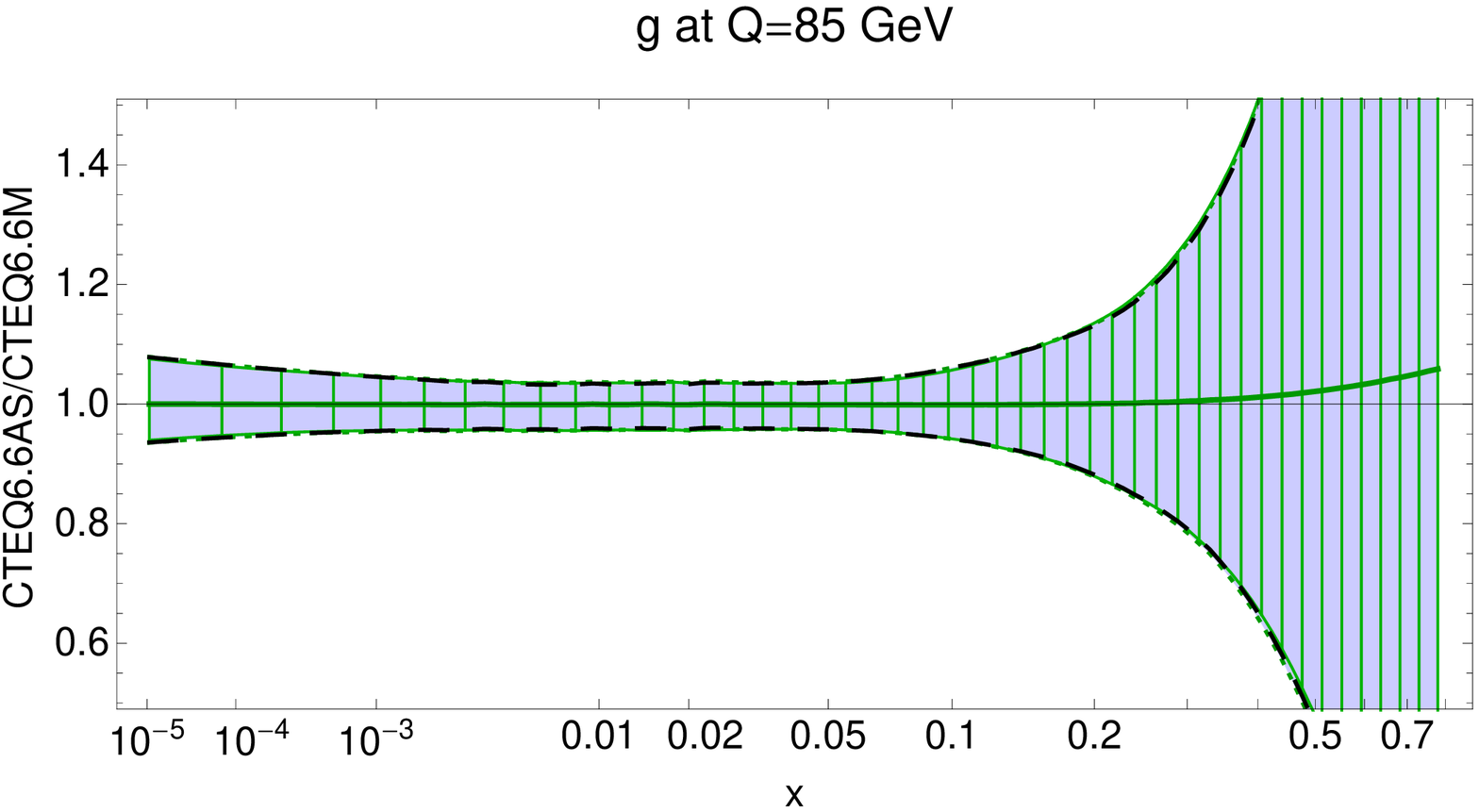}
~~\includegraphics[width=3.7in]{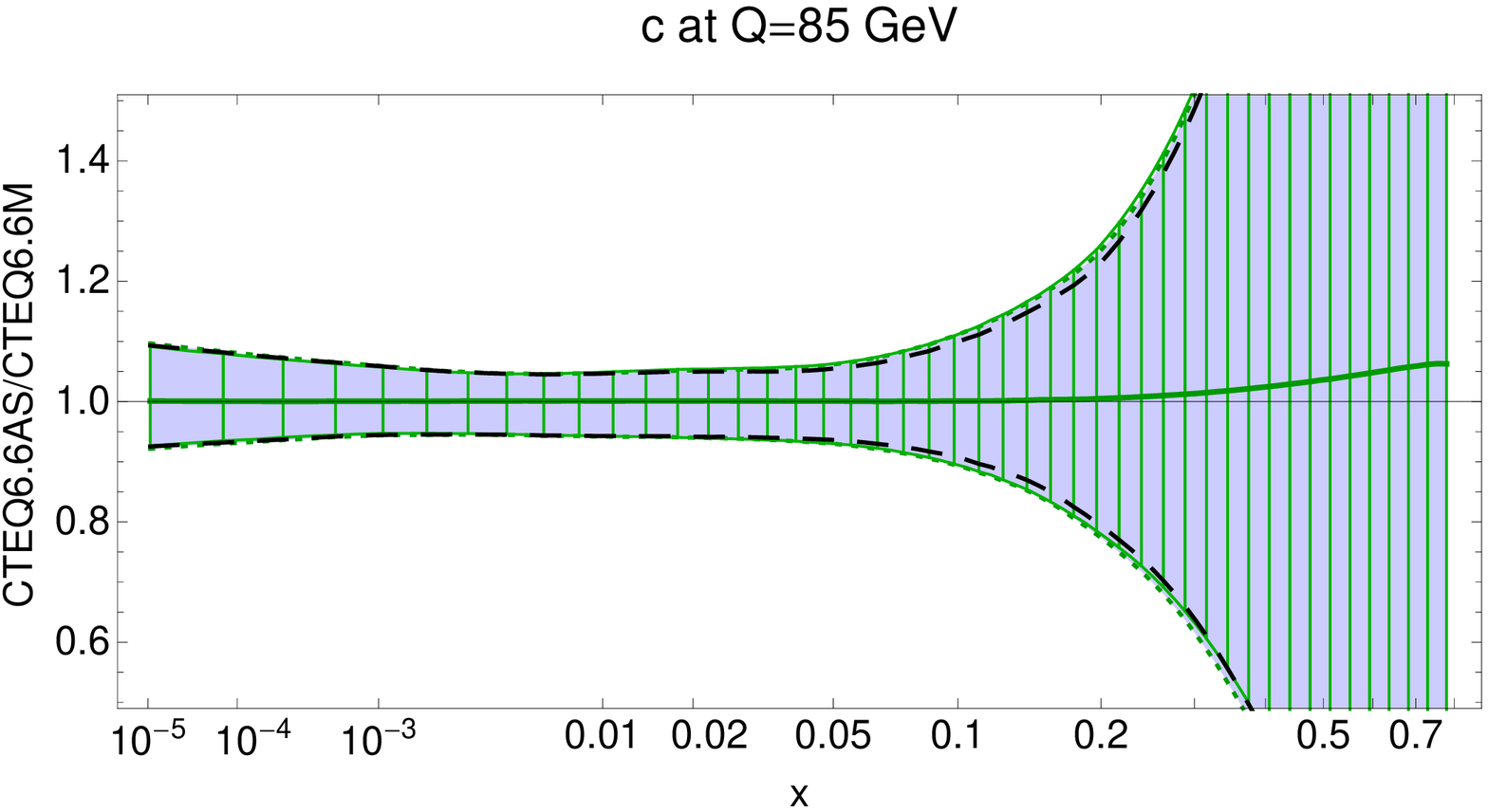}

\caption{
Comparison of CTEQ6.6+CTEQ6.6AS (blue solid) and 
CTEQ6.6FAS  (green hatched) uncertainty bands 
and CTEQ6.6 uncertainty bands
(indicated by black dashed lines), 
normalized to the standard CTEQ6.6M fit,
for gluon and charm PDFs at $Q=2$ and 85\, GeV.
\label{fig:CT66FASvsCT66}}
\end{figure}
}

\newcommand{\figPNcorrelations}
{
\begin{figure}
\begin{centering}
\includegraphics[width=3.7in]{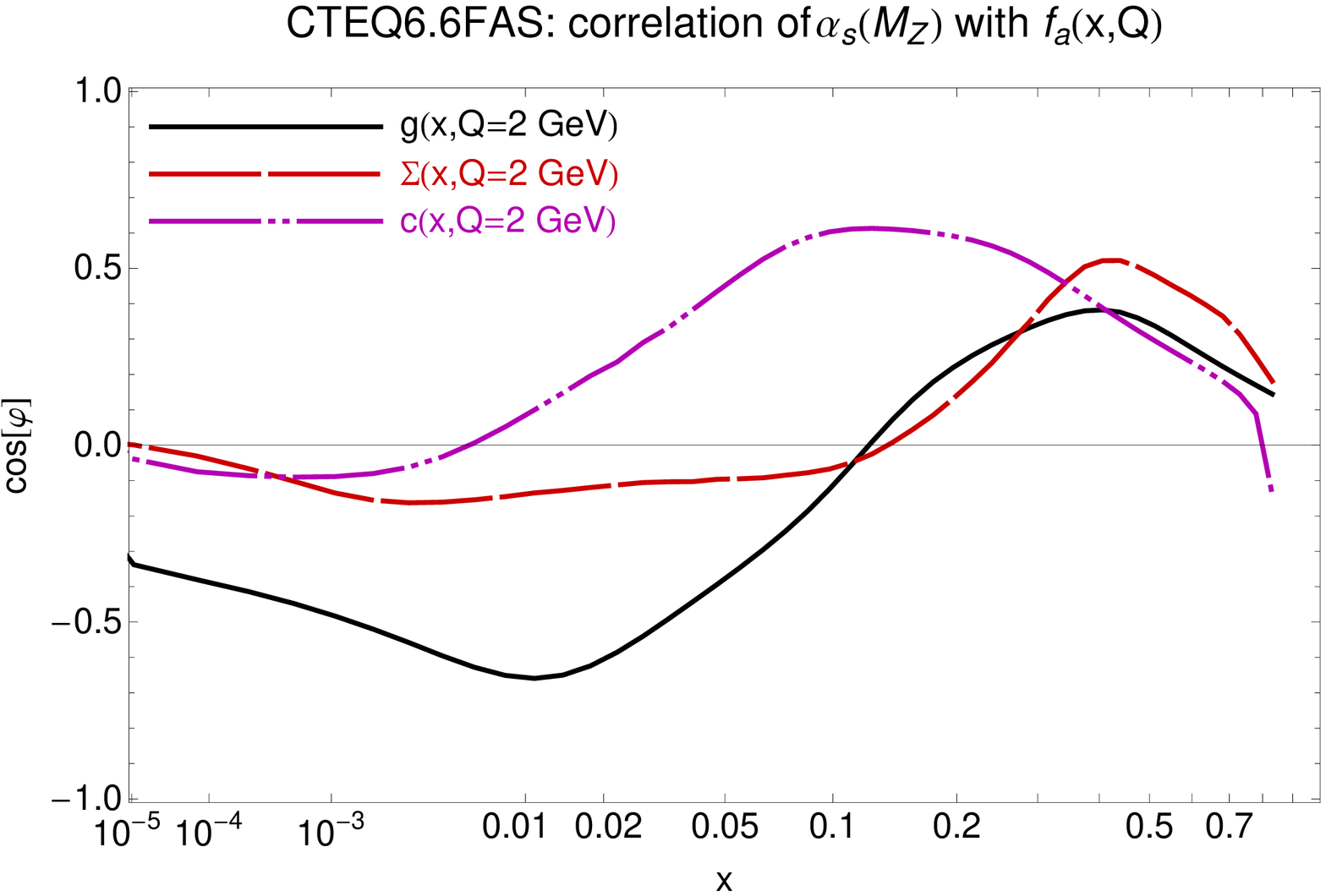}
~~\includegraphics[width=3.7in]{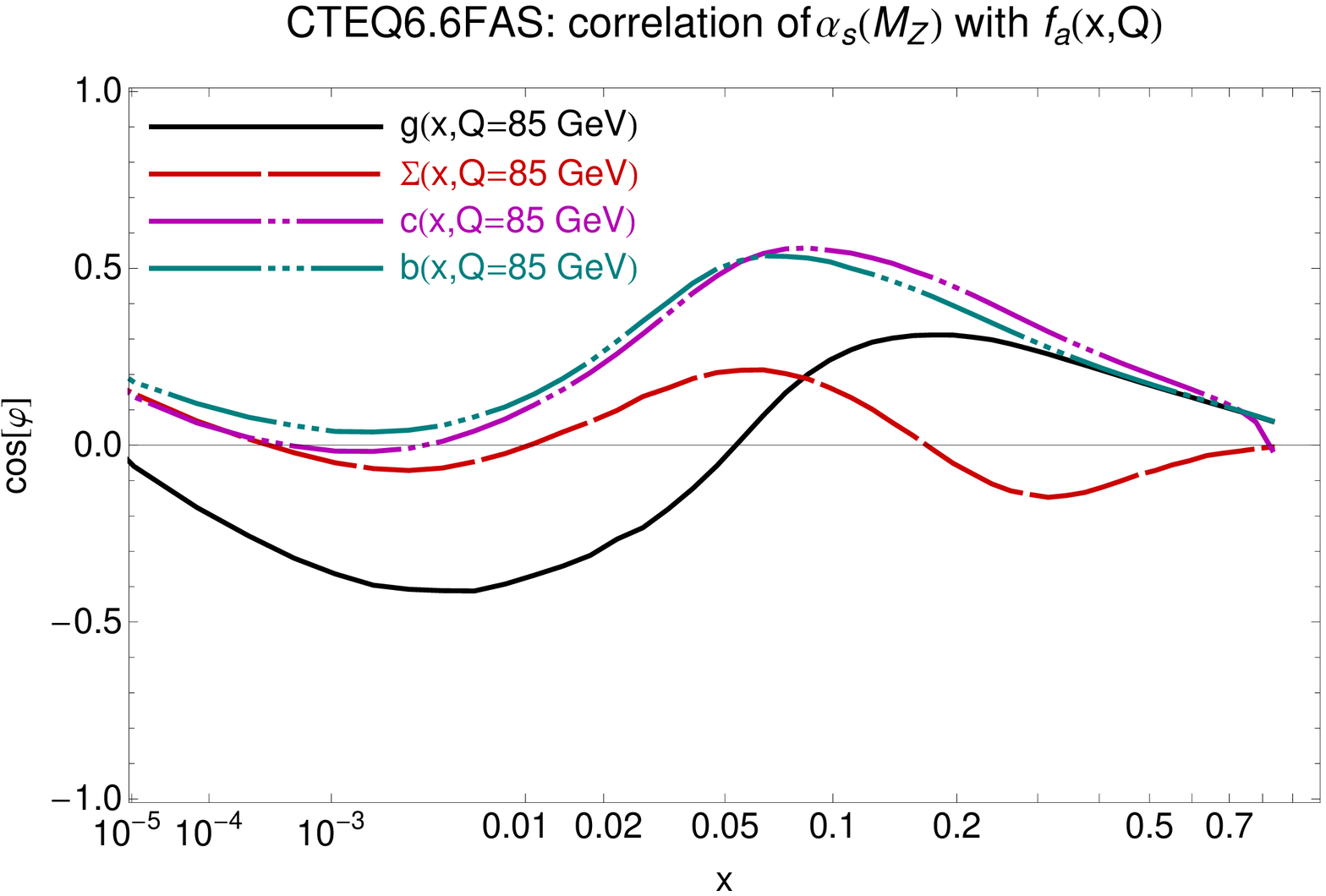}\\
\par\end{centering}

\caption{
Correlation cosine $\cos\varphi$ between the best-fit QCD coupling
$\alpha_{s}(M_{Z})$ and PDFs $f_{a}(x,Q)$, plotted as a function
of the momentum fraction $x$ at $Q=2$\, GeV
and $Q=85$\, GeV. 
\label{fig:cosphi}}
\end{figure}
}

\newcommand{\figDSDellipsesA}
{
\begin{figure}[p]
\begin{centering}
\includegraphics[width=\textwidth]{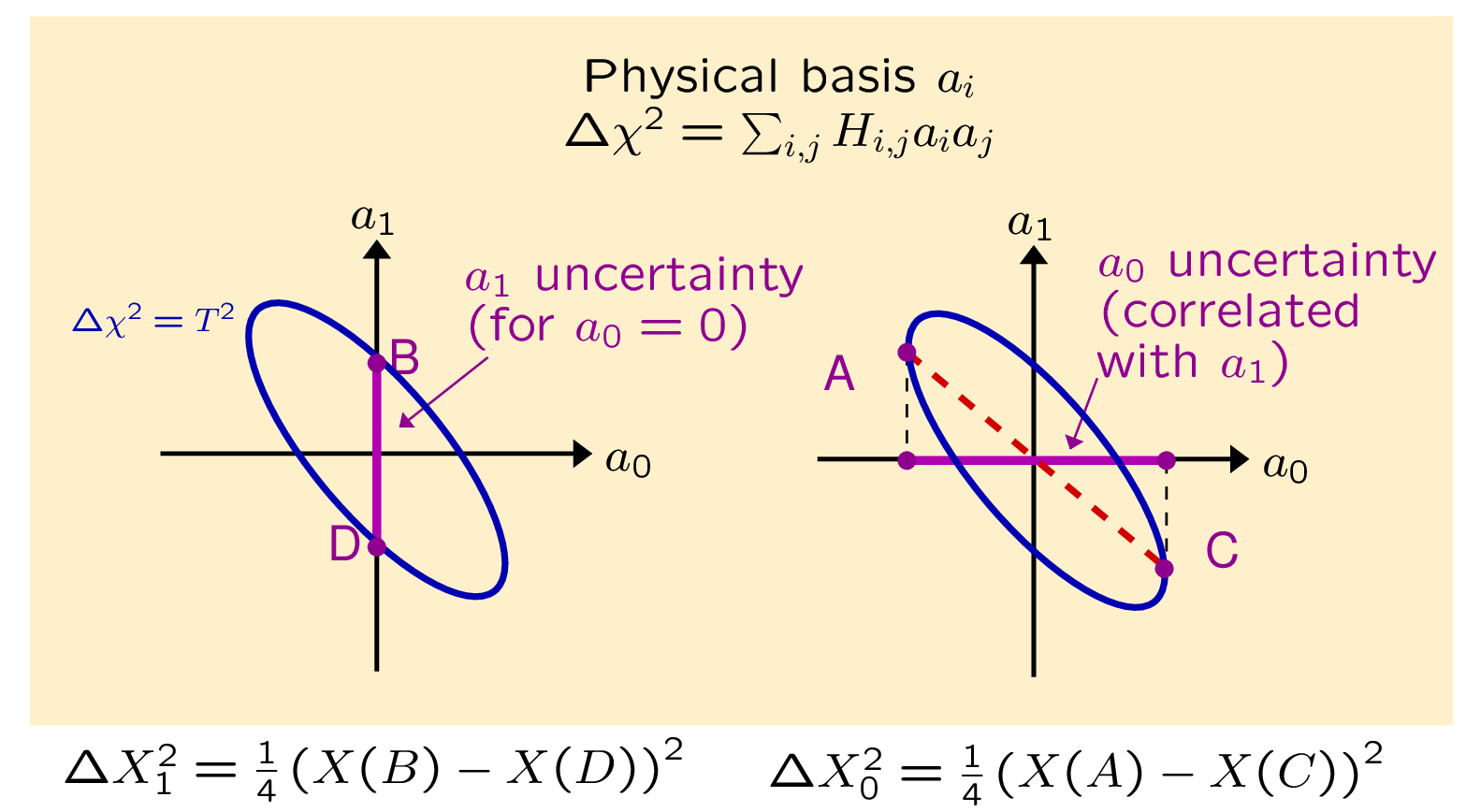}
\par\end{centering}

\caption{Imagine fitting two free parameters. The left figure shows uncertainties computed
in a fit with $a_0=0$ and free $a_1$. The right figure shows a $\chi^2$ scan over
$a_0$ with free $a_1$.
\label{fig:DSDellipsesA}}

\end{figure}
}

\newcommand{\figDSDellipsesB}
{
\begin{figure}[p]
\begin{centering}
\includegraphics[width=\textwidth]{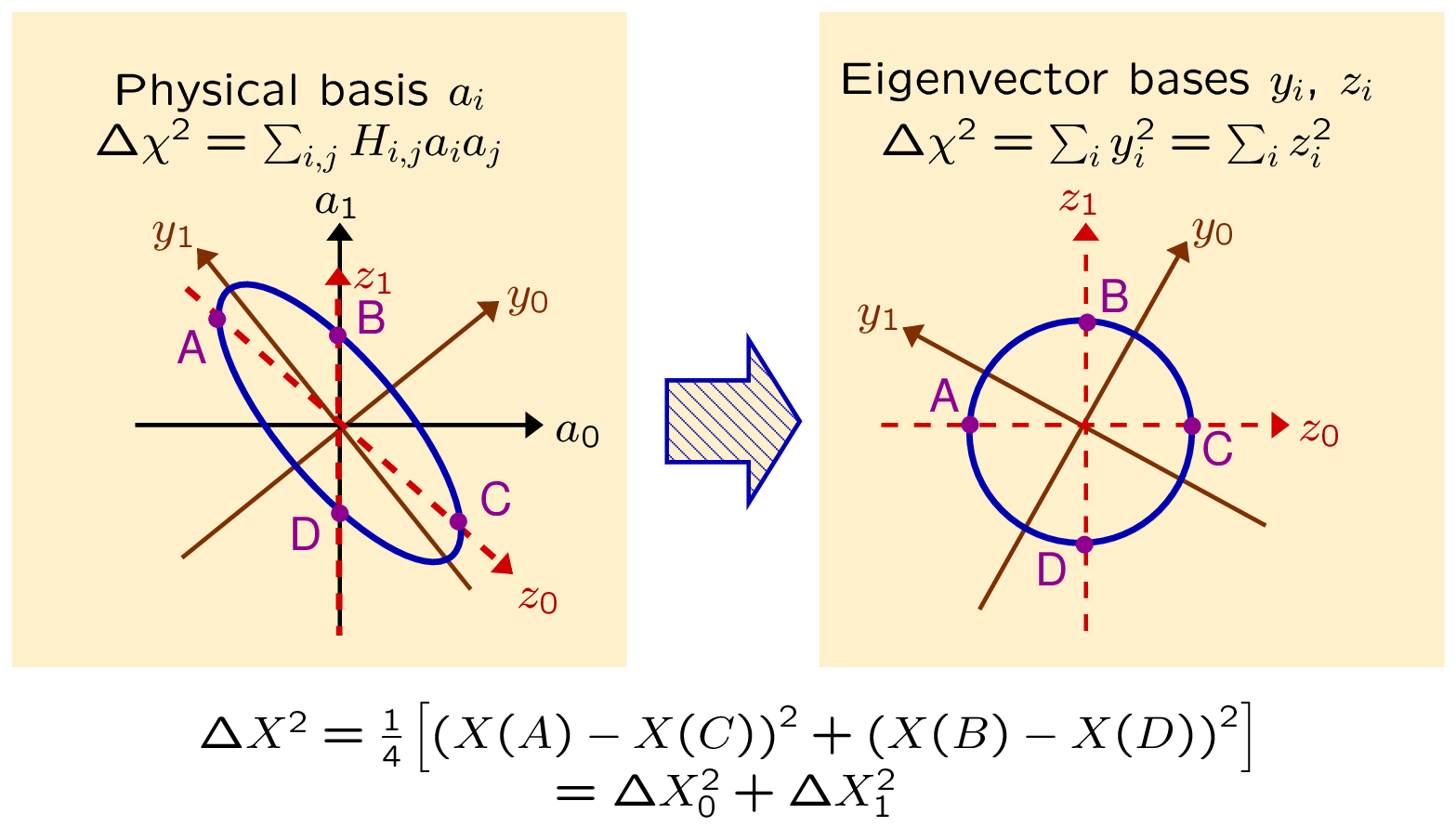}
\par\end{centering}

\caption{Imagine fitting two free parameters.
Relations are illustrated,
between the basis of physical parameters $\{a_0, a_1\}$,
the principal axis basis $\{y_0, y_1\}$, 
and the $\alpha_s$-excursion basis $\{z_0, z_1\}$.
\label{fig:DSDellipsesB}}

\end{figure}
}

%% file: CTEQ66AS-tables.tex
\newcommand{\tblcs}
{
\begin{table}
\begin{tabular}{|c||c|c|c|c||c|}

\hline
Process & \multicolumn{4}{|c||}{CTEQ6.6+CTEQ6.6AS} &
CTEQ6.6FAS \tabularnewline

\hline\hline

$t\overline{t}$ (171 GeV) & $\sigma_{0}$ & $\Delta\sigma_{PDF}$ & $\Delta\sigma_{\alpha_{s}}$ & $\Delta\sigma$ &
 $\sigma_{0}\pm\Delta\sigma$ \tabularnewline
\hline
LHC, 7 TeV & 157.41 & 10.97 & 7.54 & 13.31 & $160.10\pm13.93$\tabularnewline
\hline 
LHC, 10 TeV & 396.50 & 18.75 & 16.10 & 24.71 & $400.48\pm25.74$\tabularnewline
\hline 
LHC, 14 TeV & 877.19 & 28.79 & 30.78 & 42.15 & $881.62\pm44.27$\tabularnewline

\hline\hline

$gg\rightarrow{H}$ (120 GeV) & $\sigma_{0}$ & $\Delta\sigma_{PDF}$ & $\Delta\sigma_{\alpha_{s}}$ & $\Delta\sigma$ &
 $\sigma_{0}\pm\Delta\sigma$ \tabularnewline
\hline
Tevatron, 1.96 TeV & 0.63 & 0.042 & 0.032 & 0.053 & $0.64\pm0.055$\tabularnewline
\hline 
LHC, 7 TeV & 10.70 & 0.31 & 0.32 & 0.45 & $10.70\pm0.48$\tabularnewline
\hline 
LHC, 10 TeV & 20.33 & 0.66 & 0.56 & 0.87 & $20.28\pm0.93$\tabularnewline
\hline 
LHC, 14 TeV & 35.75 & 1.31 & 0.94 & 1.61 & $35.63\pm1.70$\tabularnewline

\hline\hline

$gg\rightarrow{H}$ (160 GeV) & $\sigma_{0}$ & $\Delta\sigma_{PDF}$ & $\Delta\sigma_{\alpha_{s}}$ & $\Delta\sigma$ &
 $\sigma_{0}\pm\Delta\sigma$ \tabularnewline
\hline
Tevatron, 1.96 TeV & 0.26 & 0.026 & 0.015 & 0.030 & $0.26\pm0.031$\tabularnewline
\hline 
LHC, 7 TeV & 5.86 & 0.16 & 0.18 & 0.24 & $5.88\pm0.26$\tabularnewline
\hline 
LHC, 10 TeV & 11.73 & 0.33 & 0.33 & 0.47 & $11.72\pm0.50$\tabularnewline
\hline 
LHC, 14 TeV & 21.48 & 0.68 & 0.56 & 0.88 & $21.43\pm0.94$\tabularnewline

\hline\hline

$gg\rightarrow{H}$ (250 GeV) & $\sigma_{0}$ & $\Delta\sigma_{PDF}$ & $\Delta\sigma_{\alpha_{s}}$ & $\Delta\sigma$ &
 $\sigma_{0}\pm\Delta\sigma$ \tabularnewline
\hline
Tevatron, 1.96 TeV & 0.055 & 0.0099 & 0.0044 & 0.011 & $0.058\pm0.012$\tabularnewline
\hline 
LHC, 7 TeV & 2.30 & 0.085 & 0.081 & 0.12 & $2.32\pm0.12$\tabularnewline
\hline 
LHC, 10 TeV & 5.08 & 0.14 & 0.15 & 0.21 & $5.10\pm0.22$\tabularnewline
\hline 
LHC, 14 TeV & 10.03 & 0.26 & 0.27 & 0.37 & $10.04\pm0.41$\tabularnewline
\hline
\end{tabular}

\caption{
Predicted cross sections for $t\overline{t}$ production
and Higgs boson production at the Tevatron and LHC, for different
scattering energies.
Cross sections are given in picobarns. The renormalization and factorization 
scales are chosen to be the top quark mass and Higgs boson mass for the 
$t \bar t$ and $gg \to H$ cross sections, respectively. The columns
present the central
prediction $\sigma_0$, PDF
uncertainty $\Delta_{PDF}$,
$\alpha_s$ uncertainty $\Delta_{\alpha_s}$, and 
total PDF+$\alpha_s$ uncertainty $\Delta\sigma$, computed according
to the CTEQ6.6AS and CTEQ6.6FAS methods. 
\label{tbl:zhaoli}}

\end{table}
}

%% file: CTEQ66AS-sec1.tex
\section{Introduction
\label{sec:Introduction}}

The global analysis of quantum chromodynamics (QCD) refers to the use 
of data from
many short-distance scattering processes to construct,
within some approximations,
universal parton distribution functions (PDFs) that can be used
to calculate hadronic cross sections for the QCD and electroweak theories.
An important part of the analysis is to determine the {\em uncertainties}
of the PDFs.
The uncertainties have several sources,
both experimental and theoretical.
An example of an uncorrelated experimental uncertainty,
i.e., an uncertainty induced by {\em data},
is the statistical error on the cross section measurements
used as the input to the global analysis.
An example of a correlated experimental uncertainty, also induced by {\em data}, is
the luminosity error, which modifies the
normalization of all cross sections measured by one experiment by
the same factor. An example of a theoretical uncertainty is the choice 
of a momentum scale $Q$ used in the theoretical calculations 
for some process; this uncertainty depends on the order of 
perturbative approximation 
(next-to-leading order, NLO, in $\alpha_{s}$
for the CTEQ analysis in this paper)
and is highly correlated between different points
in the phase space of the process.
Methods for calculating these various uncertainties have been
developed for the CTEQ analysis\,\cite{CTEQunc}
and other analyses of PDFs\,\cite{MRSTunc, MSTWforthelhc, ZEUSunc, otherunc, NeuralNetunc}.

Two overall theoretical uncertainties of the global analysis 
are the ``parametrization error'' and the uncertainty in the QCD coupling
strength $\alpha_s$.
The first concerns the parametric form for the nonperturbative PDFs
at a low momentum scale $Q_{0}$,
used as a boundary condition for predicting 
the PDFs at other scales through their perturbative evolution.
The choice of the functional form affects the predictions 
and so contributes an uncertainty to the published PDFs.

The strong coupling $\alpha_{s}$ is another basic 
parameter that affects every theoretical cross section in the global
fit. As with the PDF parametrizations, one must decide how to 
evaluate the scale dependence of the running coupling constant
$\alpha_{s}(Q)$ and provide a value of $\alpha_s$ at some scale
$Q^\prime$ as a boundary condition. Both choices contribute 
additional uncertainties to the determination of PDFs.
The $Q$ dependence of $\alpha_{s}(Q)$ is not uniquely determined,
but estimated from the NLO approximation of the renormalization group 
(RG) equation for $\alpha_{s}(Q)$:
\begin{equation}\label{eq:rge}
Q \frac{d\alpha_s}{dQ} = -\frac{\beta_{0}}{2\pi}\alpha^{2}_s
-\frac{\beta_{1}}{8\pi^{2}}\alpha^{3}_s
+O(\alpha^{4}_s),
\end{equation}
where $\beta_{0}=11-(2/3)n_{f}$, and
$\beta_{1}=102-(38/3)n_{f}$.
In NLO perturbation theory, we neglect terms of order $\alpha^{4}_s$.
Then there are different solutions of (\ref{eq:rge}),
formally equivalent in the NLO approximation, but differing
in higher orders of perturbation theory. Previous studies have shown 
that the associated uncertainty, coming from the approximation for 
the $Q$ dependence of $\alpha_{s}(Q)$, is small compared to other
sources of the PDF uncertainty \cite{CTEQ6AB}.

In this paper
we are concerned with the value of $\alpha_s$ at a typical 
scale ($Q$) for hard interactions. The differential equation 
(\ref{eq:rge}) requires an initial value, commonly taken from a
combination of precise experimental measurements.
The conventional choice is the value of $\alpha_{s}(Q)$
at the $Z$ boson mass, $Q=M_{Z}$.
Any value of $Q$ could be taken to set the initial value,
but $Q=M_{Z}$ is the natural choice, because $M_{Z}$ is known precisely,
and some experiments determine the strong coupling 
directly at this momentum scale. 

The input value of $\alpha_{s}(M_{Z})$ and its uncertainty are 
continually being refined, as new experimental data are obtained.
The global PDF analysis itself could ``determine''
(i.e., estimate within the errors) the value of $\alpha_{\rm
  s}(M_{Z})$ from the comprehensive hadronic data it examines. 
However, the resulting determination turns out to be quite uncertain: 
the included scattering processes
(deep-inelastic scattering, Drell-Yan process,
$p\overline{p} \rightarrow$ jets, etc.) have limited accuracy 
for measuring $\alpha_{s}(M_{Z})$, precisely because they depend 
on uncertain PDFs.

The most accurate value of $\alpha_{s}(M_{Z})$
is instead available from a worldwide compilation provided by the
Particle Data Group (PDG) \cite{Bethke:2009, PDG2008}.
The 2009 world average value of $\alpha_{s}$ based on eight 
different measurement techniques is \cite{Bethke:2009}
\begin{equation}\label{eq:aSbethke}
\alpha_{s}(M_{Z}) = 0.1184\ \pm\ 0.0007.
\end{equation}
Most experimental data that are an input to this world average
are independent of parton distributions.
For example, precision data from $e^{+}\,e^{-}$ annihilation
at LEP provide strong and direct constraints on $\alpha_{s}(M_{Z})$,
independently of the PDFs.
The uncertainty range in Eq.\,(\ref{eq:aSbethke})
nominally corresponds to the 68\% confidence level (CL);
yet, when assigning this level,
one must be aware that the analysis
mixes NLO and NNLO calculations in various
techniques and relies on some theoretical models.
Even with these caveats,
the accuracy of the world-average value is clearly superior
to the value determined solely from the hadronic scattering data. 

Traditionally, our analysis presents the  {\em best-fit} PDFs and
their parametrization uncertainties for a constant value 
of $\alpha_{s}(M_{Z})$ close to its latest world-average central
value. Separately, the uncertainty in the PDFs induced by 
the uncertainty in $\alpha_{s}(M_{Z})$ is assessed, 
by producing a few {\em alternative PDF fits} for a range of values 
of $\alpha_{s}(M_{Z})$ around the central $\alpha_{s}(M_{Z})$ 
value \cite{CTEQ6AB}. We add the two uncertainties in quadrature 
to estimate the total
uncertainty resulting both from the PDF
parametrization and $\alpha_{s}$.
In this paper, we examine this procedure (designated as ``fitting
method 1'') in detail, in order to determine how well it captures
the correlation, or {\em interplay}, of the $\alpha_{s}$ uncertainty and PDF uncertainty.

On the other hand, the most complete method to evaluate the
combined $\mathrm{PDF} \, + \, \alpha_s$ uncertainty in the global fit is to vary
the theoretical value of $\alpha_{s}(M_{Z})$ as an additional fitting
parameter,
while {\em including} the world-average value of $\alpha_{s}(M_{Z})$
with its experimental uncertainty as a precise experimental constraint
on $\alpha_{s}(M_{Z})$,
in combination with the rest of the hadronic data.
The uncertainty from $\alpha_{s}$ can then be deduced by this technique
-- designated as ``fitting method 2'' -- on the same footing, 
and by the same Hessian techniques, as the uncertainties 
in the PDF parameters.\cite{CTEQunc}

Our main finding is that the simpler framework of the
fitting method 1 (based on the $\alpha_{s}$ series) is sufficient to
reproduce {\em all} correlated dependence of $\alpha_{s}$ and
PDF parameters arising in the fuller treatment of the fitting method 2.
Consequently, the $\alpha_{s}$ series of CTEQ6.6 fits,
generated in fitting method 1,
is sufficient for determining the total PDF+$\alpha_{s}$ uncertainty in
theory calculations, e.g., predictions for the Large Hadron Collider (LHC).
This conclusion is borne out both numerically and formally,
within the quadratic approximation for the log-likelihood 
function $\chi^2$ in the vicinity of the best fit.
The total $\mathrm{PDF} \, + \, \alpha_s$ uncertainty can be thus calculated without
resorting to the alternative, quite elaborate, 
methods~\cite{MSTWalphas,NeuralNetunc,NNPDFalphas}, 
but simply by computing the theoretical
cross sections for two additional CTEQ6.6 PDF sets
for the extreme values of $\alpha_{s}$,
besides the usual cross sections for the 44 CTEQ6.6 eigenvector PDFs. 

Our conclusion that the addition in quadrature is sufficient, 
regardless of the magnitude of the correlation, 
follows from an easily overlooked distinction between the ``absence of correlation'', 
on one hand, and ``the independent addition of the
$\alpha_s$ uncertainty and PDF uncertainty'', on the other hand.  
As we show, variations in $\alpha_s$ generally induce compensating
adjustments in the preferred PDF parameters (correlation) 
to preserve agreement with
those experimental data sets that simultaneously constrain $\alpha_s$
and the PDFs. At the same time, it is possible to
define an ``$\alpha_s$ uncertainty'' that quantifies all correlation
effects, is independent from the PDF uncertainty for the
best-fit $\alpha_s$, and can be added in quadrature.

Section~\ref{sec:Theory} presents the new $\alpha_{s}$ series of PDFs,
based on the 2009 world average value of $\alpha_{s}(M_{Z})=0.118$
and four additional PDF sets found for
$0.116 \leq \alpha_{s} \leq 0.120$.
This series, denoted as CTEQ6.6AS, 
extends the CTEQ6.6 series of PDFs \cite{cteq66};
it uses the same input data and methods as the CTEQ6.6 study. 
In the same section,
we also introduce an alternative fit, designated CTEQ6.6FAS,
with a floating $\alpha_{s}$,
constrained by the world-average value as an experimental input.
We provide a heuristic explanation of why the CTEQ6.6AS and CTEQ6.6FAS methods should result
in close estimates of the total $\mathrm{PDF} \, + \, \alpha_s$ uncertainty.
The detailed proof is given in the Appendix.
In Section\ \ref{sec:Numerical}, we compare the CTEQ6.6AS and CTEQ6.6FAS 
methods numerically. Also, applications based on the resulting PDFs
are shown, including our estimates for 
the current $\alpha_{s}$ uncertainty 
on select theoretical predictions for the LHC.

Section~\ref{sec:Correlations} addresses a separate issue, 
the degree of the correlation between the $\alpha_s$ and PDF uncertainties
imposed by the hadronic data in the current fits. We elucidate this
issue by applying the correlation analysis of
Refs.~\cite{CTEQunc,Nadolsky:2001yg,cteq66}
to explore which data sets in the CTEQ6.6 sample impose
the tightest constraints on $\alpha_{s}$. 
The paper concludes in Section \ref{sec:Conclusions}
with our recommendation to use the $\alpha_{s}$ series, CTEQ6.6AS,
combined in quadrature with CTEQ6.6,
to assess the effect of $\alpha_{s}$ uncertainty
on theoretical predictions.

%% file: CTEQ66AS-sec2.tex
\section{QCD coupling as an input and output of the PDF analysis 
\label{sec:Theory}}

\subsection{Diagonalization of the error matrix with respect to $\alpha_{s}$}
Perturbative QCD cross sections calculated to next-to-leading order 
(NLO) have uncertainties arising from
(1) the residual scale dependence of the cross section
due to the truncation of the series at NLO,
(2) the PDF uncertainties derived from the global PDF fits
and (3) the uncertainty in the value of $\alpha_{s}(M_{Z})$
used in the cross section calculation. 
The residual scale dependence will depend on the individual 
perturbative cross section being evaluated.
The PDF uncertainty can be determined
using the PDF eigenvector sets provided by the global fitting  groups.
CTEQ and MSTW provide eigenvector PDFs (44 for CTEQ6.6  \cite{cteq66} and 40 
for MSTW2008 \cite{MSTWalphas}) determined using the Hessian method,
while the NNPDF collaboration provides $\sim 1,000$\, PDFs\,\cite{NeuralNetunc}.
The value of $\alpha_{s}(M_{Z})$ used in any cross section evaluation
must be the same as its value in the global fit providing the PDFs.
Thus, the uncertainty in $\alpha_{s}(M_{Z})$
must be evaluated by using the PDF sets in which 
the same value of $\alpha_{s}(M_{Z})$ has been assumed.

The treatment of the $\alpha_{s}$ uncertainties in any NLO cross section 
evaluation depends on their relative size compared to the intrinsic 
PDF uncertainties, and on whether there exists a correlation between 
the value of $\alpha_{s}(M_{Z})$ and the Hessian error PDFs.
One cannot exclude the possibility that the PDF and $\alpha_{s}$
uncertainties affect each other. For example, it has been
known for a long time that the shape of the gluon PDF found from the
global fit changes considerably if $\alpha_{s}(M_{Z})$ is changed in the
fit. 

There are two prevailing approaches to the choice of the value of
$\alpha_{s}(M_{Z})$ in the global PDF fits:
either the world average is taken as an {\em input},
or $\alpha_{s}(M_{Z})$ is determined as an {\em output} of the 
global fit. The CTEQ, HERAPDF and NNPDF sets of PDFs use the first approach,
while MSTW uses the second. 
Related to this choice is the evaluation of the uncertainty
in $\alpha_{s}(M_{Z})$:
either the uncertainty can be directly related to that
determined by the world average,
or the uncertainty can be determined by examining the impact 
of varying $\alpha_{s}(M_{Z})$ in the global fit. 
Again, CTEQ and the NNPDF  collaborations have used
the first approach, and MSTW the second.

The reason behind the first approach is that the most
precise processes contributing to the world average ($\tau$ and
quarkonium decays and $e^+ e^-$ event shapes at the $Z$-pole energy) are
free of the PDF uncertainties and place stronger constraints 
on the value of $\alpha_{s}(M_{Z})$ than the hadronic scattering
data alone. Knowledge gained from those processes,
summarized in the most recent PDG value of $\alpha_{s}(M_{Z})$,
naturally provides a useful external input to the global QCD analysis.
After all, the idea of {\em global} analysis is that QCD
is a fundamental theory that describes all aspects of
strong interactions. The value of $\alpha_{s}(M_{Z})$ is universal.
The value that we use should be the most accurately determined value.

The second approach has the advantage of being independent from other
sources. It circumvents the issue of combining
constraints on $\alpha_{s}(M_{Z})$ from heterogeneous measurements and mixing
different theoretical frameworks in the combination procedure. 
The price to pay, however, is the loss of the constraining power 
supplied by the most precise measurements. 

As an example of the uncertainty in $\alpha_{s}(M_{Z})$,
consider again the 2009 world average \cite{Bethke:2009} in
Eq.~(\ref{eq:aSbethke}),
its determination is dominated by the processes that do not require the PDFs.
The interpretation of the confidence level assigned to this uncertainty
(nominally 68\%) is still not fully settled for the reasons mentioned
in Section \ref{sec:Introduction}.
For the purposes of this study, we adopt a
somewhat more conservative estimate of the uncertainty proposed at the
2009 Les Houches workshop\,\cite{LesHouches2009} and adopted by the 
PDF4LHC working group\,\cite{PDF4LHC},
\begin{equation}\label{eq:LHalphas}
\alpha_{s}(M_{Z}) = 0.118 \pm 0.002
\hspace{1cm} \mbox{(90\% CL)},
\end{equation}
corresponding to an uncertainty of $\pm 0.0012$ at the 68\% CL. 

Compare that to the value obtained by MSTW from 
fitting $\alpha_s$ in the PDF analysis \cite{MSTWalphas,MSTWforthelhc},
\begin{equation}\label{eq:MSTWas1}
\alpha_{s}(M_{Z}) = 0.1202 \ \ 
\left\{
\begin{array}{c}
 + 0.0012 \\
 - 0.0015 \\
\end{array}
\right.   \hspace{1cm} (68\%\ {\rm CL}),
\end{equation}

\begin{equation}\label{eq:MSTWas1p}
\alpha_{s}(M_{Z}) = 0.1202 \ \ 
\left\{
\begin{array}{c}
 + 0.0039 \\
 - 0.0034 \\
\end{array}
\right.   \hspace{1cm} (90\%\ {\rm CL})
\end{equation}
at next-to-leading order;
or
\begin{equation}\label{eq:MSTWas2}
\alpha_{s}(M_{Z}) = 0.1171 \pm 0.0014
\hspace{1cm} (68\%\ {\rm CL})
\end{equation}
at next-to-next-to-leading order.
Or, compare that to our own determination of
$\alpha_{s}(M_{Z}) = 0.118 \pm 0.005$ (90\% CL) from a fit
without the world-average constraint that is described below, 
cf.~Eq.~(\ref{eq:olderas}).
The result is consistent with that in Eq.~(\ref{eq:aSbethke});
but the uncertainty on $\alpha_{s}(M_{Z})$
is considerably larger if $\alpha_{s}$ is fitted 
without the world-average constraint. 
The central value of $\alpha_{s}(M_{Z})$ returned by this 
fit is $0.118$, which 
coincides with the fixed value used in the CTEQ6.6 PDF analysis.

One can also envision a third, most general approach,
in which the $\alpha_{s}(M_{Z})$ range 
measured by the most precise techniques, i.e. the world average value, is 
included as an {\em  input}
together with the usual hadronic scattering data;
and the theoretical value of $\alpha_{s}(M_{Z})$, fitted 
as a free parameter together with the PDF parameters, 
is returned as an {\em output} constrained by the
combination of {\em all} measurements.
Such a fit is the most direct in quantifying improvements
in the accuracy of the output $\alpha_{s}(M_{Z})$
resulting from the hadronic scattering data (as compared to the more
precise constraints); as well as in probing the correlation between
$\alpha_{s}(M_{Z})$ and PDF parameters.

In this paper, we examine the CTEQ global hadronic data and
the world-average $\alpha_s(M_Z)$ in Eq.~(\ref{eq:LHalphas}) 
according to this more general approach,  and apply the 
usual Hessian technique \cite{CTEQunc} to study 
the combined $\mathrm{PDF} \, + \, \alpha_s$  uncertainty (i.e. the usual PDF shape parameters plus 1 more parameter for $\alpha_s$).
However, when implemented straightforwardly, this approach runs into a practical
inconvenience: each of the eigenvector PDFs is associated 
with its own value of $\alpha_s(M_Z)$.

The quadratic approximation
provides a remarkable bypass for this shortcoming.
By applying the Data Set Diagonalization method 
introduced in Ref.~\cite{DSD}, it is
possible to re-diagonalize the parameter space 
so that only one eigenvector corresponds to the change of $\alpha_s$, 
whereas all the other eigenvectors are immune to the change in
$\alpha_s$.
(As shown in Ref.~\cite{DSD}, this conclusion also holds 
for any other distinct parameter in the global analysis.)
In other words, it is easy to construct a pair of additional PDF sets
corresponding to the maximal tolerated excursions of $\alpha_{s}$.
The $\alpha_{s}$ uncertainty computed from the difference
of these PDFs can be added in quadrature to the CTEQ6.6
uncertainty to reproduce the combined uncertainty
with the full PDF-$\alpha_{s}$ correlation.
We refer to the reduced computation as ``fitting method 1,''
and the full computation as ``fitting method 2.''   

\figDSDellipsesA
\figDSDellipsesB

A formal proof that methods 1 and 2 are equivalent,
within the quadratic approximation,
is given in the Appendix.
Here we provide a heuristic argument that illustrates
the choice of the two PDF eigenvector
sets probing the $\alpha_{s}$ uncertainty.

The $\alpha_{s}$ parameter in the full fit
is associated with a new twenty-third direction
that is orthogonal to the hyperplane of the PDF parameters 
for the central value of $\alpha_{s}(M_{Z})=0.118$;
{\it i.e.}, the 22-parameter space probed by the CTEQ6.6 PDF set.
Originally, the orthonormal eigenvector PDFs
of the CTEQ6.6 fit correspond to a representation 
such that the excursion by the same distance from the best fit 
in the 22-dimensional
hyperspace results in the same increase of $\chi^2$;
that is, the surfaces of constant $\chi^2$
are 22-dimensional hyperspheres.

If $\alpha_{s}$ is allowed to deviate from its best-fit value,
a preferred direction emerges in the 22-dimensional hyperspace,
along which $\Delta \chi^2$ grows most slowly, because changes in 
$\alpha_{s}$ are compensated by changes in 
some linear combination of the PDF parameters.
Rotate the CTEQ6.6 basis so that the ``physical'' PDF parameter $a_1$
corresponds to the variation of this preferred combination from 
its best-fit value, and analogously, 
the "physical" parameter $a_0$ corresponds to the variation of $\alpha_s(M_Z)$ from its best-fit value (which remains at $0.118$). With this choice, $a_0=0$ and $a_1=0$ at the best fit. 
All correlations between $\alpha_{s}$
and PDFs are encapsulated in the $\chi^2$ dependence
on $a_0$ and $a_1$.
Let us focus on this dependence. 
Uncertainties due to the other (combinations of) PDF parameters
$a_2,$ ..., $a_{22}$ are independent of $\alpha_{s}$ and can be
added in quadrature at the end. 

In the $\{a_0, a_1\}$ plane, the set of the allowed PDFs corresponds to
the inside of an ellipse $\Delta \chi^2 \leq T^2$, where $T$ is the
tolerance parameter or some other parameter defining the typical
$\chi^2$ at the boundary of the allowed region. The CTEQ6.6 fit 
is equivalent to probing the inside
of the ellipse for a fixed $a_0=0$, as shown in the left
inset of Fig.\,\ref{fig:DSDellipsesA}.
The contribution of the $a_1$ direction to the CTEQ6.6 uncertainty
for an observable $X$ can be evaluated as 
\begin{equation}
\Delta X_1^2 = \frac{1}{4}\left(X(B)-X(D)\right)^2,
\end{equation}
where $X(B)$ and $X(D)$ are the values of $X$ at points {\em B}
and {\em D} in the figure. 

Next, consider the minimal and maximal excursions of the parameter 
$a_0$ allowed inside the ellipse, for a free parameter $a_1$.
These excursions are reached at points {\em A} and {\em C}
shown in the right inset of  Fig.\,\ref{fig:DSDellipsesA}.
The uncertainty along this direction is 
\begin{equation}
\Delta X_0^2 = \frac{1}{4}\left(X(A)-X(C)\right)^2.
\end{equation}

In practice, points {\em A} and {\em C} are
found by a scan of the dependence of $\Delta \chi^2(a_0, a_1)$ on $a_0$, 
for a varying $a_1$.
To describe the uncertainty of an observable $X$, we 
rotate the $\{a_0,a_1\}$ basis to a basis
$\{y_0,y_1\}$ of eigenvectors of the Hessian matrix. The orthogonal directions
$\mathbf {y}_0$ and $\mathbf{y}_1$ 
specify the principal axes of the ellipse,
as seen in the left inset of Fig.\,\ref{fig:DSDellipsesB}.
Since the maximal range of the $a_0$ parameter is specified by the 
line segment $AC$ in Fig.\,\ref{fig:DSDellipsesA}
or Fig.\,\ref{fig:DSDellipsesB}, we could further rotate the
$\{y_0,y_1\}$ basis to another basis 
$\{z_0,z_1\}$, after rescaling the 
eigenvectors $\mathbf{y}_0$ and $\mathbf{y}_1$ to unity, as shown in the  
right inset of Fig.\,\ref{fig:DSDellipsesB}. 
The vector $\mathbf{z}_0$ is chosen to be along the {\em AC} direction.
It is shown in the Appendix that the vector  $\mathbf{z}_1$ 
is along the direction of the 
line segment {\em BD}.  It is perpendicular to the $\mathbf{z}_0$ direction 
in the $\{z_0,z_1\}$ representation, even though the  $\mathbf{z}_0$ and  
$\mathbf{z}_1$ directions 
are not perpendicular in the $\{a_0,a_1\}$ representation.
Taking this conclusion for granted here, we could 
easily see from the  left inset of Fig.\,\ref{fig:DSDellipsesB} that 
there is no $a_0$ dependence along the $\mathbf{z}_1$ direction. 
From the right inset, 
the total uncertainty of the observable $X$ can be computed as 
\begin{eqnarray}
\Delta X^2 &=&
\frac{1}{4}\left[\left(X(A)-X(C)\right)^2+\left(X(B)-X(D)\right)^2\right]\\
&=&\Delta X_0^2 + \Delta X_1^2.
\end{eqnarray}
This  uncertainty is equal to the quadrature sum of the uncertainties 
along the {\em AC} and {\em BD} directions, as has been stated.

We see that the $\mathbf z_0$ and $\mathbf z_1$ directions, which are initially not orthogonal in the $\{a_0, a_1\}$ basis, can be made such by rotation and scaling.
Alternatively, the orthonormal $\{z_0, z_1\}$ basis can be obtained from the 
$\{a_0, a_1\}$ basis by a shear transformation in the negative $a_1$ direction,
followed by a scaling transformation along the $a_0$ direction.
In the 23-dimensional case,
the condition $a_0=z_0=0$ defines the
hyperplane spanned by the 22 CTEQ6.6 PDF parameters, for the fixed
best-fit $\alpha_{s}$.
This hyperplane is made orthogonal to $z_0$ by a rotation and scaling
transformation, or, equivalently, by a shear and scaling
transformation, analogously to the 2-dimensional case.
See the Appendix for the full discussion.

\subsection{Fits with a variable $\alpha_{s}$: explicit realizations}

We will now explicitly construct two next-to-leading order (NLO) 
fits of the kinds described above, 
using the CTEQ6.6 sample\,\cite{cteq66} of hadronic data~\footnote{Like in all CTEQ global fits, we include
full NLO matrix elements in DIS and most Drell-Yan observables. In calculations for
inclusive jet production and  W lepton asymmetry, where the full NLO/resummed results
are prohibitively CPU-extensive,  K-factor tables are used to look up the ratio of the NLO
cross section to the LO cross section (with the LO cross section
calculated using the NLO PDFs, and using the 2-loop $\alpha_s$) separately for each data point.
The look-up tables depend very weakly on the input PDF parameters. They are updated in the
course of the fitting to ensure that the K-factors have not drifted from
their initial values, and we again check  that the calculations retain their full
NLO accuracy at the end of the global fit.}. 
To this end, we modify the setup of the CTEQ6.6 analysis 
to allow $\alpha_{s}(M_{Z})$ to vary within the global fit,
and to constrain these variations by
the world-average (w.a.) value $\left( 
\alpha_{s}\right)_{w.a.} \pm \left(\delta\alpha_{s}\right)_{w.a.} 
= 0.118 \pm 0.002$ (at 90\% CL) included as a separate data input,
in addition to the complete CTEQ6.6 set of 
hadronic scattering data.
Agreement with this {\em precision data value} is just as desirable (or more so)
as agreement with individual data points in the hadronic data sets.
So, to assure this agreement, we add a new contribution 
$\chi_{\alpha_{s}}^{2}$
to the global log-likelihood function in the global analysis:
\begin{equation}\label{eq:quadrature}
\chi^{2}=\chi_{\mbox{CTEQ6.6}}^{2}+\chi_{\alpha_{s}}^{2},
\end{equation}
where
\begin{equation}
\chi_{\alpha_{s}}^{2} = \lambda\left[
\frac{\alpha_{s}(M_{z})-\left(\alpha_{s}(M_{z})\right)_{w.a.}}
{\left(\delta\alpha_{s}\right)_{w.a.}}\right]^{2}.
\end{equation}
The $\chi_{\alpha_{s}}^{2}$ term is multiplied by a weighting factor
$\lambda$ to match the confidence interval of the world-average $\alpha_{s}$
with the tolerance on the increase in $\chi^{2}$ allowed for acceptable
fits. We choose $\lambda$ so that $\alpha_{s}(M_{z})$ values
outside of the 90\% CL interval, $\alpha_{s}(M_{Z})\leq0.116$ or
$\alpha_{s}(M_{Z})\geq0.120$,
result in a penalty beyond the tolerance for the increase in $\chi^2$.

Without the contribution of the world-average value,
the constraints of the global fit on $\alpha_{s}$ are relatively weak,
\begin{equation}\label{eq:olderas}
\alpha_{s}(M_{Z})=0.118 \pm 0.005 \hspace{1cm} \mbox{(90\% CL)}.
\end{equation}
The central value returned by this fit is practically identical
to either the PDG or Les Houches workshop central values. 
When the world-average constraint is included, the final result of the
global analysis changes to
\begin{equation}\label{alphasCT66FAS}
\alpha_{s}(M_{Z})=0.1180 \pm 0.0019 \hspace{1cm} \mbox{(90\% CL)}.
\end{equation}
Again, its central value is practically the same, 
but the uncertainty is much smaller.
Thus the constraint on $\alpha_{s}(M_{Z})$
is dominated by the world-average uncertainty,
$(\delta\alpha_{s})_{w.a.}$.

We explore the vicinity of the best fit in two ways.  
First, we construct best-fit PDFs sets 
for four alternative values of $\alpha_{s}(M_{Z})$, 
\begin{equation}\label{eq:5values}
\alpha_{s}(M_{Z})=0.116,\ \ 0.117,\ \ 0.119,\mbox{ and } \ 0.120.
\end{equation}
These PDFs are named as
\begin{equation}
{\rm CTEQ6.6AS} = AS_{-2},\ \ AS_{-1},\ \ AS_{+1},\mbox{ and } \ AS_{+2}.
\end{equation}

The CTEQ6.6AS PDFs for the two extreme variations,
$AS_{-2}$ and $AS_{+2}$,
correspond to slightly more than two standard deviations according to
the PDG error in Eq.~(\ref{eq:aSbethke}), or approximately to 
the 90\% CL uncertainty according to the 2009 Les Houches
prescription in Eq.~(\ref{eq:LHalphas}).
The intermediate PDF sets, $AS_{-1}$ and $AS_{+1}$,
provide additional information on the $\alpha_{s}$ dependence.

Alternatively, the combined PDF and $\alpha_{s}$ uncertainty
is quantified by the diagonalization of the Hessian matrix,
in terms of 46 extreme PDF eigenvector sets
for 23 independent combinations of theoretical parameters.
Each eigenvector set is associated with its own $\alpha_{s}(M_{Z})$ value
within the best-fit range of Eq.~(\ref{alphasCT66FAS}). This series of eigenvector 
PDFs is called CTEQ6.6FAS. It will be compared with the CTEQ6.6AS series 
in the next section.

%% file: CTEQ66AS-sec3.tex
\section{Numerical results \label{sec:Numerical}}

\subsection{Comparison of the PDF and $\alpha_{s}$ uncertainties}

The  $AS_{-2}$ and $AS_{+2}$ PDFs of the CTEQ6.6AS series 
for the gluon, $u$ quark, and $s$ quark are compared with 
the CTEQ6.6 PDF uncertainty band
for the momentum scale $Q=2$ GeV in Fig.\,\ref{fig:GUSlowQ},
and for $Q=85$\, GeV in Fig.\,\ref{fig:GUShighQ}.
Each figure shows the ratio of $f(x,Q)$ in the $AS_n$ set to the 
CTEQ6.6M PDF of the same flavor (solid and dashed curves), 
as well as the asymmetric CTEQ6.6 PDF uncertainty for this flavor (shaded
region), as functions of the momentum fraction $x$. 
The upper and lower boundaries $1\pm [\Delta f(x,Q)]_{\pm}/f(x,Q)$ of
the PDF uncertainty band are given by the asymmetric PDF errors, as
\begin{eqnarray}{\label{eq:band}}
\left[\Delta{f}(x,Q)\right]_{+}
&=& \sqrt{\sum_{i=1}^{22} \left[ f_{i}(x,Q)-f_{0}(x,Q) \right]^{2}} 
{\rm ~~for~~} f_{i}>f_{0}, \nonumber \\
\left[\Delta{f}(x,Q)\right]_{-}
&=& \sqrt{ \sum_{i=1}^{22} \left[ f_{i}(x,Q)-f_{0}(x,Q) \right]^{2}} 
{\rm ~~for~~} f_{i}<f_{0}.
\end{eqnarray}

For the shown partons ($g$, $u$, and $s$),
the PDF uncertainty is larger than the $\alpha_{s}$ dependence
over the range $0.116 \leq \alpha_{s}(M_{Z}) \leq 0.120$.
The figures for the $d$ quark (not shown) looks similar to the 
figures for the $u$ quark. The relation between the PDF and
$\alpha_s$ uncertainties for $\bar u$ and $\bar d$ quarks 
is in between these relations for the $u$ and $s$ quarks.
Finally, the $c$ and $b$ uncertainties are qualitatively similar 
to those for $g$.

\figGUSlowQ

\figGUShighQ

\figPNbands

We now add the CTEQ6.6 PDF uncertainty in quadrature to the CTEQ6.6AS
$\alpha_{s}$ uncertainty, and compare this sum to the uncertainty of
the 23-parameter fit CTEQ6.6FAS with the floating $\alpha_{s}$ and the world
average constraint.
The result is shown in Fig.\,\ref{fig:CT66FASvsCT66}.
The figure compares the PDF uncertainty bands from the two
realizations of the $\alpha_{s}$ fit, 
for the parton flavors that show the largest $\alpha_{s}$ dependence:
gluons and charm quarks.
The CTEQ6.6 error band is indicated by dashed lines;
while the two realizations of the $\alpha_{s}$ fit by a
hatched error band with solid borders and a filled error band with
dotted borders.
We observe only small differences between the CTEQ6.6+CTEQ6.6AS
and CTEQ6.6FAS PDFs for the flavors shown in the figure. 
The differences are even smaller for the other quark PDFs.
Thus, the two realizations of the $\alpha_{s}$ series
produce nearly identical results, confirming the adequacy of adding 
the PDF and $\alpha_{s}$ uncertainties in quadrature.

\subsection{Uncertainties of cross section predictions}

Using the $\alpha_{s}$ series of PDFs, and including the PDF uncertainty,
we can estimate the uncertainties of cross section calculations.

For any calculated quantity $\sigma$,
we denote the central prediction, corresponding to $\alpha_s(M_Z)=0.118$,
by $\sigma_{0}$.
There are two contributions to the uncertainty.
The symmetric PDF uncertainty, denoted $\Delta\sigma_{\rm PDF}$,
is calculated from the Hessian error PDFs by the
``master formula''\ \cite{CTEQunc}
\begin{equation}\label{eq:PDFunc}
\Delta\sigma_{\rm PDF}=
\frac{1}{2}\sqrt{
\sum_{i=1}^{d}\left(\sigma_{i}^{(+)}-\sigma_{i}^{(-)}\right)^{2}.
}
\end{equation}
Here $d$ is the dimension of the parameter space,
and $\sigma_{i}^{(\pm)}$ is calculated with the eigenvector PDFs.\footnote{%
For these PDFs, based on CTEQ6.6,
the number of fitting parameters is $d = 22$.
PDF sets $+i$ and $-i$ are variations of the central fit corresponding to 
displacements in the 
$+$ and $-$ directions along eigenvector $i$.
The symmetric error formula, Eq.~(\ref{eq:PDFunc}), is sufficient for 
the comparison here; at other times, we would use 
more detailed asymmetric error estimates, Eq.~(\ref{eq:band}).}
The $\alpha_{s}$ uncertainty of $\sigma$ is
\begin{equation}\label{eq:ASunc}
\Delta\sigma_{\alpha_{s}}=\frac{1}{2}
\sqrt{\left[\sigma_{0}(A_{-2})-\sigma_{0}(A_{2})\right]^{2}}.
\end{equation}
The combined uncertainty $\Delta{\sigma}$ for CTEQ6.6+CTEQ6.6AS is
\begin{equation}\label{eq:combined}
\left(\Delta{\sigma}\right)^{2}
=\left(\Delta{\sigma}_{\rm PDF}\right)^2
+\left(\Delta{\sigma}_{\alpha_{s}}\right)^{2}.
\end{equation}
For the CTEQ6.6FAS series,
the full uncertainty is computed according to Eq.~(\ref{eq:PDFunc}) for $d=23$.
Then the prediction for the quantity $\sigma$ is
\begin{equation}
\label{eq:pred}
\sigma = \sigma_{0} \pm \Delta{\sigma}.
\end{equation}

\tblcs
\figDRSttblhc

Table \ref{tbl:zhaoli} lists the predictions
for a sample set of NLO cross sections at the Tevatron and LHC energies.
Four processes are calculated: $t\overline{t}$
production~\cite{NLOttbarNason,NLOttbarBeenakker}, 
using a program from Ref.\,\cite{NLOttbarBeenakker};
and Standard Model Higgs boson production, 
$gg\rightarrow H$, 
for Higgs masses $M_{H}=120, 160$ and $250$\,GeV \cite{Spira:1995mt}.
The production cross sections are predicted 
for three LHC energies: 7, 10 and 14 TeV.
The central predictions $\sigma_{0}$ 
and uncertainty ranges $\Delta{\sigma}$ are given
first for CTEQ6.6+AS (fitting method 1) 
and then for CTEQ6.6FAS (fitting method 2).
Clearly, the predictions 
by the two methods are very close, even though not identical due to 
secondary effects like deviations from the quadratic approximation.

\figDRSgghlhc

In fitting method 1, we can also compare the relative sizes of 
the PDF uncertainty and the $\alpha_{s}$ uncertainty. 
For processes dominated by the gluon scattering, 
such as the $t\bar t$ production or Higgs production examined here, 
the $\alpha_s$ and PDF uncertainties can be comparable, as is observed 
in the Table. These results are illustrated for $t\bar t$ production  
at the LHC with energies 7 TeV and 14 TeV in Fig.~\ref{fig:DRSttblhc}, 
and for Higgs boson production ($M_{H}=120$\,GeV) 
at the LHC with energies 7 TeV and 14 TeV in Fig.~\ref{fig:DRSgghlhc}.
The figures show the cross sections 
from individual CTEQ6.6(AS) eigenvector sets, 
as well as the resulting PDF and $\alpha_s$ uncertainties, versus 
the corresponding $\alpha_s(M_Z)$ values.
The overall prediction based on the full CTEQ6.6+AS eigenvector set,
$\sigma \pm \Delta \sigma$, is shown by an error bar:
the inner bar is the PDF error alone; the outer bar
is the combined $\mathrm{PDF} \, + \, \alpha_s$  error. According to
the figures, the $\alpha_s$ uncertainty of the total cross section  
in the shown processes
constitutes between 70\% and 110\% of the PDF uncertainty. 

%% file: CTEQ66AS-sec4.tex
\newpage
\section{Correlation between $\alpha_s$ and the PDFs
\label{sec:Correlations} }

\figPNcorrelations

The independence of the $\alpha_s$ uncertainty from the PDF uncertainty in the
CTEQ6.6AS method does not preclude existence of some correlation
between the $\alpha_s$ and PDF parameters. This correlation arises 
from the hadronic scattering experiments, which probe a variety 
of combinations of the PDFs and $\alpha_s$. Let us now examine 
which PDF flavors are most affected by variations in $\alpha_{\rm s}(M_{Z})$, 
and which scattering experiments impose the most relevant constraints. 

For this purpose, we go back to the full
fit CTEQ6.6FAS with the floating $\alpha_s$  and 
examine the correlation between $\alpha_{\rm s}(M_{Z})$
and individual PDF $f_{a}(x,Q)$ using the method outlined
in Refs.~\cite{CTEQunc,cteq66,Nadolsky:2001yg}.
Let $Y_{i}^{(\pm)}$ (for $i=1,...,23$) denote one of 46 eigenvector 
PDFs $f_{a}(x,Q)$ for a chosen $a,$ $x,$ and $Q$, 
corresponding to the maximal acceptable displacements of orthonormal
PDF parameters $\{z_{1},...,z_{23}\}$ from their best-fit values
in the positive $(+)$ and negative $(-)$ directions, respectively.
Given 46 values of $\alpha_{\rm s}(M_{Z})$ (denoted as $X_{i}^{(\pm)}$),  
we compute the correlation
cosine,
\begin{equation}\label{cosphi}
\cos\varphi=\frac{1}{4\Delta{X}\,\Delta{Y}}
\sum_{i=1}^{23}\left(X_{i}^{(+)}-X_{i}^{(-)}\right)
\left(Y_{i}^{(+)}-Y_{i}^{(-)}\right),
\end{equation}
where $\Delta{X}$  and $\Delta{Y}$ are the symmetric PDF errors,
found from Eq.~(\ref{eq:PDFunc}) for $d=23$.
Values of $\cos\varphi$ close to $+1$, 0, $-1$ indicate strong
correlation, no correlation,
and strong anti-correlation between $\alpha_{\rm s}(M_{Z})$
and the PDF $f_{a}(x,Q)$ in question. 

Fig.~\ref{fig:cosphi} shows $\cos\varphi$ versus $x$,
for the PDFs that have the largest correlations with
$\alpha_{\rm s}(M_{Z})$, at $Q=2$ and 85 GeV.
Other PDFs have little correlation with $\alpha_{\rm s}(M_{Z})$
and are not shown in the figure.
We also identify the experimental data sets that 
cause the largest correlations.
As shown in Fig.~\ref{fig:cosphi}, at $Q=2$~GeV ,
the most significant correlation (or anticorrelation) of
$\alpha_s(M_Z)$ occurs with: 
\begin{itemize}
\item{the gluon PDF $g(x,Q)$
(anticorrelated at $x\sim0.01$ due to neutral-current
DIS constraints from HERA);}
\item{the singlet PDF $\Sigma(x,Q)$
(correlated at $x\sim0.4$ due to constraints
imposed by BCDMS and NMC neutral-current deep-inelastic scattering (DIS) data);}
\item{the heavy-quark PDFs, $c(x,Q)$ and $b(x,Q)$
(correlated at $x=0.05-0.2$ due to constraints by HERA charm and bottom
semi-inclusive DIS data).}
\end{itemize}
At $Q=85$ GeV, these correlations are reduced by the PDF evolution,
with the exception of the heavy-quark PDFs. 
The same conclusions can be drawn from the 
simpler CTEQ6.6AS analysis, as well as from the 
correlation analysis by NNPDF \cite{NNPDFalphas}.

The best-fit value of $\alpha_{\rm s}(M_{Z})$ is thus determined by several
types of the data, probing the gluon evolution in DIS at moderately small
$x,$ the singlet PDF evolution in DIS at large $x$, and HERA charm 
semi-inclusive DIS data.
 The correlation of each kind disappears if the relevant
data set is removed. For example, it is believed that the low-$Q$/large-$x$
BCDMS and NMC DIS data prefer somewhat lower $\alpha_{\rm s}(M_{Z})$
values than the rest of the experiments \cite{alphasBCDMS},
indicating the possible presence of higher-twist terms
that are not explicitly included in most PDF analyses
\cite{BCDMShighertwist}.
With the most suspect part of these data excluded,
the central $\alpha_{\rm s}(M_{Z})$ value indeed increases to 0.119 -- 0.120;
the spike in $\cos\varphi$ for the singlet PDF $\Sigma(x,Q)$
at $x\sim 0.4$ also disappears from Fig.\,\ref{fig:cosphi}.

%% file: CTEQ66AS-sec5.tex
\section{Conclusion
\label{sec:Conclusions}}

We conclude with this important point:
while there are correlations between
$\alpha_{\rm s}(M_{Z})$ and some PDFs,
as demonstrated in Figure \ref{fig:cosphi},
the total uncertainty of the CTEQ6.6FAS fit is essentially the same 
as the CTEQ6.6+CTEQ6.6AS fit,
as demonstrated in Figure \ref{fig:CT66FASvsCT66}. The 
CTEQ6.6+CTEQ6.6AS method
captures all the correlation of $\alpha_{\rm s}$ and PDF parameters as a mathematical
consequence of the quadratic approximation, upon which the Hessian
method is based. The ability of the quadrature method to reproduce the
total PDF+$\alpha_s$ uncertainty (also observed in a related NNPDF
study \cite{NNPDFalphas}) 
is thus more than a coincidence.

We may predict cross sections using the CTEQ6.6FAS PDFs,
with their $23\times 2$ Hessian eigenvector PDF sets.
The results are shown in the final column of Table \ref{tbl:zhaoli}.
The predictions agree well with those of the simpler method based on
the CTEQ6.6+CTEQ6.6AS PDF sets.
This again justifies our recommendation to use the CTEQ6.6+CTEQ6.6AS PDFs
to estimate uncertainties induced by the PDFs and $\alpha_s$,
according to Eq.\,(\ref{eq:combined}).

The experimental measurements included in the world average give a smaller
$\alpha_{\rm s}$ uncertainty than the hadronic scattering data can match.
Therefore, when making predictions for the LHC or Tevatron colliders,
the most realistic estimate of the $\alpha_{\rm s}$ uncertainty
will result from the CTEQ6.6+CTEQ6.6AS $\alpha_{\rm s}$ series,
which covers
\begin{equation}
\alpha_{\rm s}(M_{Z}) = 0.116,\ 0.117,\ 0.118,\ 0.119,\mbox{ and } 0.120.
\end{equation}
The central value of this series (assumed in the CTEQ6.6 eigenvector set)  
is closest to the world average.
The PDFs for $\alpha_s$ values at 0.116 and 0.120 provide an estimate of the
$\alpha_s$ uncertainty at approximately 90\% CL, which is to be
combined in quadrature with the CTEQ6.6 PDF uncertainty. The CTEQ6.6AS PDF
sets are available from the CTEQ6.6 PDF website \cite{CT66website} and as
a part of the LHAPDF library \cite{LHAPDF}.

\quad\\
{\bf Acknowledgments}\\
We thank S. Forte, J. Rojo, R. Thorne, A. Vicini, 
participants of the PDF4LHC and Les Houches 2009 workshops, 
and members of CTEQ for stimulating discussions.
This work was supported in part
by the U.S.DOE Early Career Research Award DE-SC0003870;
by the U.S. National Science Foundation under grant PHY-0855561;
by the National Science Council of Taiwan under grants
NSC-98-2112-M-133-002-MY3 and NSC-99-2918-I-133-001;
by LHC Theory Initiative Travel Fellowship awarded by the U.S. National Science Foundation under grant PHY-0705862;
and by Lightner-Sams Foundation.
C.-P. Y. would also like to thank the hospitality of
National Center for Theoretical Sciences in Taiwan and
Center for High Energy Physics, Peking University, in China,
where part of this work was done.

%% file: CTEQ66AS-app1.tex
\appendix

\section{Error matrix diagonalization for the QCD coupling
\label{app:DSDalphas}}
In this Appendix we provide a proof of the method advocated in this study,
whereby an uncertainty induced by 
the small variation of $\alpha_s$ value is added 
in quadrature with the fixed-$\alpha_s$ PDF uncertainty.
The only necessary assumptions are the usual quadratic
approximation for $\chi^2$, and the understanding that
the PDF sets for the ``up'' and ``down'' variations of $\alpha_s$  
are obtained through fits in which all of the PDF parameters are free.

%Like the Hessian method in general, the theorem relies on the assumption that 
%$\chi^2$ can be approximated by a quadratic function of the fitting parameters.
Assume we have an initial fit in which there are $N$ free parameters, with 
an additional parameter held fixed.  The additional parameter can be any 
additional degree of freedom that one wishes to study. Here it is intended to 
be $\alpha_s(M_Z)$; but no special features of that parameter will be invoked.

We begin by carrying out the traditional Hessian diagonalization
procedure, applicable 
in the vicinity of the best-fit (BF) combination of the PDF parameters.
Starting from the ``shape'' parameters that control the input PDFs at 
scale $Q_{0}$, we define new fitting parameters $a_1,\dots,a_N$ that are 
coefficients of the eigenvectors of the original Hessian matrix, to obtain
\begin{equation}
\chi^2  \, =  \, \chi_{\mathrm{BF}}^2  \, +  \, \sum_{i=1}^N a_i^{\, 2}\; .
\label{eq:origchisq}
\end{equation}

We now wish to include an additional degree of freedom in the fit. 
Hence we introduce a new parameter $a_0$, which could be defined as proportional 
to the deviation of $\alpha_s(M_Z)$ from its best-fit value, 
 $\alpha_s(M_Z) -  \alpha_{s, \mathrm{BF}}(M_Z)$; or better, 
as proportional to $\ln[\alpha_s(M_Z) /  \alpha_{s, \mathrm{BF}}(M_Z)]$.
Including the new degree of freedom, we have, in general, 
\begin{equation}
\chi^2 \, = \, \chi_{\mathrm{BF}}^2  \, + \, b \, a_0 \, + \, 
\sum_{i=0}^N \sum_{j=0}^N H_{ij} \, a_i \, a_j \;,
\label{eq:chisq}
\end{equation}
where
\begin{equation}
\mathbf{H}  \, = \, 
\begin{pmatrix} 
  1 & p_1 & p_2 & p_3 & \dots & p_N \\ 
p_1 &   1 &   0 & 0   & \dots &   0 \\ 
p_2 &   0 &   1 & 0   & \dots &   0 \\ 
p_3 &   0 &   0 & 1   & \dots &   0 \\ 
\cdot & \cdot & \cdot & \cdot & \dots & \cdot \\
\cdot & \cdot & \cdot & \cdot & \dots & \cdot \\
\cdot & \cdot & \cdot & \cdot & \dots & \cdot \\
p_N &   0 &   0 & 0   & \dots &   1  \\
\end{pmatrix} \; .
\label{eq:newH}
\end{equation}
Here
\begin{equation}
p_{i}=\frac{1}{2} 
\left( \frac{\partial^{2}\chi^{2}}
{\partial{a_{0}}\partial{a_{i}}}
\right)_{\rm BF}.
\end{equation}
$\mathbf{H}$ describes the correlation between $a_{0}$ and $a_{i}$,
as it gives the variation of $\chi^{2}$ in the $\{a_{0},a_{i}\}$-subspace.
The linear term $b \, a_0$ in (\ref{eq:chisq})
allows for the possibility that the new minimum might not be at $a_0 = 0$.
Note that
$ b =\left( \partial\chi^{2}/\partial{a}_{0} \right)_{\rm BF}$,
which describes how $\chi^{2}$ varies with $a_{0}$
at the original best fit.
The new diagonal element $H_{00}$ was chosen to be 1 by including an 
appropriate scaling factor in the definition of $a_0$.

%pn Apr 22
As $a_0$ varies, the PDF parameters $a_1,...,a_N$ are adjusted so
that to reduce the increase in $\chi^2$. The direction along which
this increase is minimal (direction $AC$ in
Fig.\,\ref{fig:DSDellipsesA}) corresponds to $\partial \chi^2/\partial a_i=0$
for $i=1,...N$, {\it i.e.}, it is a line given by $a_i=-p_i a_0$
according to Eqs.~(\ref{eq:chisq}) and (\ref{eq:newH}). 

Define new variables $z_i$ as
\begin{eqnarray}
z_i &=& a_i + p_i\, a_0 \, , \mbox{~~~for }i=1,...,N,\label{ziai}\\
z_0 &=& \sqrt{1-C^2} \, a_0 + \frac{b}{2\sqrt{1-C^2}}, \label{z0a0}
\end{eqnarray}
where \begin{equation}\label{eq:Cdef}
C = \sqrt{\sum_{i=1}^{N} p^{2}_{i}}.
\end{equation}
Since $\chi^2$ is assumed to have a minimum, hence, $0 \leq C <1$. The transformation (\ref{ziai}, \ref{z0a0}) 
to the $z_i$ coordinates consists of an $a_0$-dependent translation
(shear) along the PDF coordinates
$a_i$, followed by a scaling and a translation along the $\alpha_s$
coordinate $a_0$. 
In the new coordinates, $\chi^2$ is diagonal, 
\begin{equation}
\chi^2 \, = \, \chi_{\mathrm{BF}}^2  \, - \frac{b^2}{4 (1-C^2)} \, + \, \sum_{i=0}^N z^2_i .
\label{eq:chisqz}
\end{equation}
Variations along $z_0$ and $z_i$ for $i=1,...N$ are explicitly
independent, and we can calculate the symmetric uncertainty 
$\Delta X$ for an observable $X(z_0,z_1,...,z_N)$ by 
\begin{equation}
(\Delta X)^2 \, = \, \frac{1}{4}\sum_{i=0}^N\left[X(0,...,z_i=T,...,0)
- X(0,...,z_i=-T,...,0)\right]^2, 
\label{eq:DelFsq}
\end{equation}
for tolerance $\Delta \chi^2 = T^2$.

Below, we will focus on the case that $b=0$, which is valid in this paper. 
From  Eqs.\  (\ref{ziai}) and (\ref{z0a0}),
the extreme values of $a_i$ consistent with  $\Delta \chi^2 = T^2$ are
$a_i^\pm =z_i^\pm =\pm T$. They occur at $a_0=z_0=0$ ({\it i.e.}, at the
best-fit $\alpha_s$) and $z_k = 0$ for $k\neq
i$.  The extreme values of $a_0$ occur at $z_0^\pm =\pm T$ and
$z_i=0$ for
$i=1,...,N$, {\it i.e.,} $a_0^\pm = \pm T/\sqrt{1-C^2}$ 
 and $a_i^\pm = - p_i a_0^\pm$.
Hence Eq.\ (\ref{eq:DelFsq}) corresponds exactly 
to adding in quadrature the uncertainty based on the allowed range of $a_i$ computed 
at $a_0 = 0$ (analogous to the CTEQ6.6 uncertainty 
in our application) to the uncertainty based 
on the allowed range of $a_0$, computed with $a_i$ at their preferred
values for the extreme excursions in $a_0$ (analogous to the CTEQ6.6AS
uncertainty).

An alternative way to construct the $z_i$ basis is to find the
eigenvectors of the $(N+1)\times(N+1)$ Hessian matrix 
$\mathbf{H}$ in Eq.\ (\ref{eq:newH}) and then perform an additional
orthogonal transformation.
To find the eigenvectors,
note that $\mathbf{H}$ can be written in a block form
\begin{equation}
\mathbf{H}= \left(
\begin{array}{cc}
 1 & \mathbf{p}^{T} \\ \mathbf{p} & \mathbf{I} \end{array}
\right)
\end{equation}
where 1 is a unit element,
$\mathbf{p}^T$ is a row vector $(p_{1}, p_{2}, ..., p_{N})$,
and $\mathbf{I}$ is an $N\times N$ unit matrix.

\begin{itemize}
\item{One eigenvector is 
\begin{equation}
\mathbf{V}^{(0)} = F \left( 
\begin{array}{c}
C \\ \mathbf{p}
\end{array}
\right),
\end{equation}
corresponding to an eigenvalue $\lambda_{0}=1+C$.
The normalization factor is $F=1/(\sqrt{2}C)$.
}
\item{A second eigenvector is
\begin{equation}
\mathbf{V}^{(1)}=F\left(
\begin{array}{c}
-C \\ \mathbf{p}
\end{array}
\right),
\end{equation}
with an eigenvalue $\lambda_{1}=1-C$.
}
\item{Now consider vectors of the form
\begin{equation}
\mathbf{V} = \left(
\begin{array}{c}
0 \\ \mathbf{q}
\end{array}
\right)
\end{equation}
where $\mathbf{p}^{T}\mathbf{q}=0$.
These are orthogonal to the first two eigenvectors;
the $N$-dimensional vector $\mathbf{q}$ is orthogonal to ${\mathbf p}$.
For any ${\mathbf q}$, we have $\mathbf{H\,V = V}$.
All their eigenvalues are $\lambda_{j}=1$, for $j=2, 3, \dots , N$.
}
\end{itemize}

So we have a set of eigenvectors of $\mathbf{H}$,
denoted by
$\mathbf{V}^{(0)}, \mathbf{V}^{(1)}, \mathbf{V}^{(2)}, 
\dots , \mathbf{V}^{(N)}$, and
satisfying the orthonormality condition
\begin{equation}\label{eq:orthonormal}
\sum_{k=0}^{N} V^{(i)}_{k}V^{(j)}_{k}=\delta_{ij} .
\end{equation}
The next step is to use the  eigenvectors
$\mathbf{V}^{(j)}$ to define new coordinates 
$t_0,\dots,t_N$ by
\begin{equation}
 a_i \, = \, \sum_{j=0}^N \, t_j \, V_i^{(j)} \, .
\end{equation}
It is easy to compute $\chi^2$ from Eq.~(\ref{eq:chisq}) by making use of
the eigenvector properties
\begin{equation}
 \mathbf{H} \, \mathbf{V}^{(j)} \, = \, \lambda_j \, \mathbf{V}^{(j)} \, ,
\end{equation}
where $\lambda_0 = 1+C$, $\lambda_1 = 1-C$, and 
$\lambda_j = 1$ for $j = 2,\dots, N$; and then using the orthogonality 
property (\ref{eq:orthonormal}).
The result has the expected diagonal form
\begin{equation}
\chi^2 \, = \, \chi_{\mathrm{BF}}^2  
\, + \,  (1+C) t_0^{\,2}
\, + \,  (1-C) t_1^{\,2}
\, + \,  \sum_{i=2}^N t_i^{\, 2}\; ,
\end{equation}
which is further simplified to
\begin{equation}
\chi^2 \, = \, \chi_{\mathrm{BF}}^2  \, 
+ \,  \sum_{i=0}^N y_i^{\, 2} \, 
\end{equation}
by rescaling of the first two coordinates: 
\begin{equation}
y_0 = \sqrt{1+C} \, t_0;\quad y_1 = \sqrt{1-C} \, t_1; \quad 
y_i = t_i\mbox{ for }i=2,\dots,N\; .
\end{equation}

The function $\chi^2$ is explicitly diagonal in the new coordinates $y_i$, but
not in the parameter $a_0$ that 
is of interest here, given that
\begin{equation}
 a_0  \, = \, y_0 / \sqrt{2(1+C)}  \, - \, y_1 / \sqrt{2(1-C)} \; .
\end{equation}
It can be made so by a further orthogonal transformation 
called the Data Set Diagonalization \cite{DSD}.
%pn It is one variable, and not a data set, that is diagonalized here.  
The general method for finding the necessary transformation is to write 
$ a_0^{\, 2} \, = \, \sum_{i=0}^N \sum_{j=0}^N G_{ij} \, w_i \, w_j \,$ and use 
the eigenvectors of $\mathbf{G}$ as new basis vectors.  However, in the present case,
the required transformation can be found more simply by observing 
that one of the new coordinates must be proportional to $a_0$, and the others must 
be orthogonal to it.  The result is 
\begin{eqnarray}
z_0 \, &=& \, {\scriptstyle\sqrt{\frac{1-C}{2}}} \, y_0 \, - \, 
{\scriptstyle \sqrt{\frac{1+C}{2}}} \, y_1 \nonumber , \\
z_1 \, &=& \, {\scriptstyle \sqrt{\frac{1+C}{2}}} \, y_0 \, + \, 
{\scriptstyle \sqrt{\frac{1-C}{2}}} \, y_1 \nonumber, \\
z_j \, &=& \; y_j \quad \mbox{for $j=2,\dots,N$} \; ,
\label{eq:ytrans}
\end{eqnarray}
which yields
\begin{equation}
\chi^{2} = \chi^{2}_{\rm BF} 
+ \sum_{j=0}^{N} z_{j}^{2},
\end{equation}
equivalent to Eq.~(\ref{eq:chisqz}) with $b=0$.
Because $\chi^{2}$ is a sum of independent terms for each $z_{i}$,
the uncertainties of these variables should be combined in quadrature.
This conclusion does not depend on the value of $C$, {\em i.e.}, the strength of correlation between 
the new degree of freedom $a_0$ and the original parameter space of $a_i$ in Eq.(\ref{eq:origchisq}). 
The dependence of $\chi^2$ on $a_0$ comes entirely from the coordinate $z_0$.
The space spanned by $z_1,\dots,z_N$ at $z_0 = 0$ has $a_0 = 0$, so it is the 
same fitting space spanned by $a_1,\dots,a_N$ in the original $N$-parameter 
fit.  Hence the uncertainty associated with $z_1,\dots,z_N$ is the same 
as the uncertainty associated with $a_1,\dots,a_N$.
It is therefore not necessary to explicitly carry out the
transformations, and the overall formula for the uncertainty is 
the result we have used in Eq.\ (\ref{eq:combined}).